\documentclass{aa}  
\usepackage{graphicx}
\usepackage{txfonts}
\begin{document}
   \title{Simulations of polarized dust emission}


   \author{V.-M. Pelkonen
          \inst{1}
          \and
          M. Juvela\inst{2}
	  \and
	  P. Padoan \inst{3} 
          }


   \institute{Observatory, University of Helsinki,
              T\"ahtitorninm\"aki, P.O.Box 14,
	      SF-00014 University of Helsinki, Finland \\
	      \email{vpelkone@astro.helsinki.fi}
	     \and
	     Department of Astronomy, University of Helsinki,
	     T\"ahtitorninm\"aki, P.O.Box 14,
	     SF-00014 University of Helsinki, Finland \\
	     \email{mika.juvela@helsinki.fi}
	     \and
	     Department of Physics, University of California, San Diego,
	     CASS/UCSD 0424, 9500 Gilman Drive, La Jolla, CA 92093-0424 \\
	     \email{ppadoan@ucsd.edu}}


   \date{}

 
  \abstract
   {}
   {The aim is to study the polarization of thermal dust emission to see if
   the alignment of grain by radiative torques could explain the observed
   relation between the degree of polarization and intensity. Predictions are
   made for polarimetry observations with the Planck satellite.}
   {Our results are based on model clouds derived from MHD
   simulations of magnetized turbulent flows. The continuum radiative transfer
   problem is solved with Monte Carlo methods in order to estimate the
   three-dimensional distribution of dust emission and the radiation field
   strength affecting the grain alignment. The minimum grain size aligned by
   the radiative torques is calculated, and the Rayleigh polarization
   reduction factor, $R$, is derived for different grain size distributions.
   We show maps of polarized thermal dust emission that are predicted by the
   models. The relation between the intensity and polarization degree is
   examined in self-gravitating cores. Furthermore, we study the effects of
   wavelength, resolution, and observational noise.}
   {We are able to reproduce the P/I-relation with the grain alignment by radiative
   torques. The decrease in intrinsic polarization and total emission means
   that sub-mm polarimetry carries only little information about the magnetic
   fields in dense cores with high visual extinction. The interpretation of
   the observations will be further complicated by the unknown magnetic field
   geometry and the fact that what is observed as cores may, in fact, be a
   superposition of several density enhancements.
   According to our calculations, Planck will be able to map dust 
   polarization reliably when $A_{\rm V}$ exceeds $\sim 2^{\rm m}$ at spatial resolution of $\sim 15'$.}
   {}


   \maketitle
%

\section{Introduction}

Magnetic fields play an important role in the dynamics of molecular clouds and
in the star formation process. In protostellar cores the topology of the magnetic fields 
can be predicted with models of core formation and evolution. However, observations of
the magnetic field structure is needed to test such models. The
best technique for studying magnetic fields in molecular clouds is to observe
the polarization of the light from background stars and the polarization of
the thermal dust emission, both arising because dust grains are aligned by
the magnetic field.

Although the polarization of starlight by aligned interstellar dust grains was
discovered over half a century ago (Hall \cite{hall}, Hiltner \cite{hiltner}),
the processes responsible for the alignment are still under study. In a recent
review by Lazarian (\cite{lazarian}) the various proposed mechanisms are
discussed extensively.

One of the first processes to be suggested was the paramagnetic mechanism
(Davis \& Greenstein \cite{davisgreen}), which is based on direct interaction
of rotating grains with the interstellar magnetic field. While this mechanism
requires stronger magnetic fields than have been uncovered by other means, it
may contribute at very small grain sizes. Purcell (\cite{purcell}) introduced
several processes to make grains very fast ``suprathermal'' rotators and
identified one of them, catalytic sites ejecting ${\rm H_{2}}$ as the major
cause of fast grain rotation. However, it couldn't explain, for example, why
the observations indicated that small grains are less aligned than large ones.

Fairly recently, the Purcell mechanism was re-evaluated in a series of papers
by Lazarian and colleagues, pointing out processes such as internal grain
wobbling (e.g. Jones \& Spitzer \cite{josp67}; Lazarian \cite{la94}; Lazarian
\& Roberge \cite{laro97}), grain flipping (Lazarian \& Draine \cite{ladr99a})
and ``nuclear relaxation'' (Lazarian \& Draine \cite{ladr99b}). All these
contribute to making the Purcell mechanism inefficient.

The radiative torque mechanism was introduced by Dolginov (\cite{do72}) and
Dolginov \& Mytrophanov (\cite{domy76}). Draine \& Weingartner (\cite{drwe96})
demonstrated the efficiency of radiative torques using numerical simulations,
while Abbas et al. (\cite{abal04}) demonstrated it in a laboratory setup. The
radiative torques use the interaction of radiation with a grain to spin it up.
The predictions of the radiative torque mechanism are roughly consistent with
observations (e.g. Lazarian et al. \cite{laal97}; Hildebrand et al.
\cite{hial00}).

In view of this, the radiative torque mechanism is a strong candidate as the
primary mechanism for grain alignment inside molecular clouds. Cho \& Lazarian
(\cite{cholazarian}) showed that even deep inside GMCs ($A_V \leq 10$) large
grains can be aligned by radiative torques, using a simple, radially symmetric
core model. Thus, far-infrared and submillimeter polarimetry could reliably
reflect the structure of magnetic fields deep inside molecular clouds.

Padoan et al. (\cite{paal01}) used numerical MHD simulations to model a
molecular cloud containing protostellar cores, and computed dust continuum
polarization for self-gravitating cores. They made the simplifying assumption
that the temperature of the dust is uniform, which allowed them to calculate
the polarization based on gas density and magnetic field vectors. They
calculated the polarization maps for two cases: one where the grain alignment
mechanism efficiency would be independent of the visual extinction and another
where it was assumed that the mechanism would stop working at a sharp cut-off
of $A_{\rm V} = 3^{\rm m}$, and thus dense cores would not contribute to the
polarization. The latter case reproduced the observed results of decreasing
polarization degree with increasing intensity in the cores, which led them to
conclude that the grains at high visual extinction appear not to be aligned. 
This would mean that submillimeter polarization maps carry little
information on the magnetic fields inside protostellar cores at visual
extinction larger than $A_{\rm V} \approx 3^{\rm m}$.

Like Padoan et al. (\cite{paal01}), we base our study on three-dimensional
simulations of magnetized gas flow, but we refine the analysis method used to
predict the polarized dust emission. Radiative transfer calculations are used
to estimate the actual three-dimensional distribution of dust emission. This
naturally reduces the weight of high column density regions, because there the
emission will be reduced by the attenuation of the heating radiation field.
Similarly, radiative transfer modelling allows us to estimate, at every point
of the cloud, the radiative torque efficiency of Cho \& Lazarian
(\cite{cholazarian}). This way we can test the hypothesis, that reduced
polarization of dense cores could be due to the non-alignment of smaller
grains. We study the general morphology of the resulting dust
polarization maps. In particular, we look at the relation between total
and polarized dust emission in dense cores. Finally, some predictions are made
concerning the dust polarization that will be observable with the Planck
satellite.

%


\section{Polarization}
\subsection{Rayleigh Polarization Reduction Factor}

\begin{figure}
\centering 
\includegraphics[width=8cm]{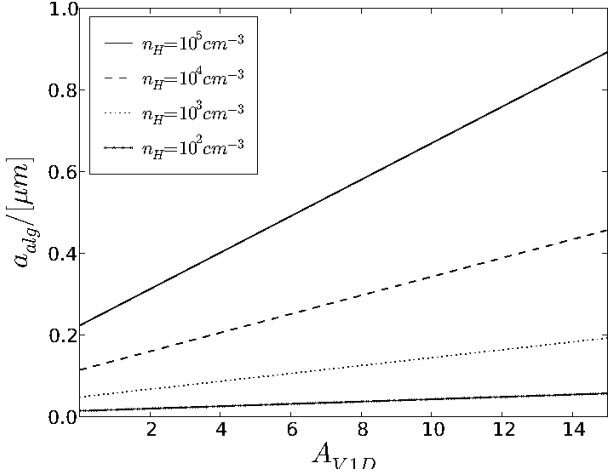} 
\caption{
Minimum aligned grain size as a function of $A_{\rm V, 1D}$ using the Weingartner
\& Draine (\cite{wedr01}) dust model and Li \& Draine (\cite{Li2001}) grain size distribution. 
} \label{ld_aalg}
\end{figure}

The Rayleigh polarization reduction factor $R$ is a measure of imperfect
alignment of the dust grains with respect to the magnetic field (Greenberg
\cite{greenberg}; see also Lee \& Draine \cite{leedraine}). The degree of
polarization is decreased, when part of the grains are misaligned with respect
to the magnetic field. In our case, small grains are not aligned by radiative
torques while the larger ones are. Cho and Lazarian (\cite{cholazarian}) give
a fitting formula for the minimum aligned grain size in the case of a spherically symmetric dark cloud:
\begin{eqnarray}
a_{{\rm alg}} = (log \;n_{\rm H})^3(A_{\rm V, 1D} + 5) / 2800 \; \mu{\rm m},
\label{eqaalg}
\end{eqnarray}
where $n_{\rm H}$ is the hydrogen density and $A_{\rm V, 1D}$ is the visual extinction used by Cho and Lazarian (\cite{cholazarian}). They used the radiation field given in Mathis et al. (\cite{maal83}), which was given as a function of the visual extinction measured along a perpendicular sightline from the surface of an opaque (in the centre $A_{\rm V} \sim 200^{\rm m}$), spherical cloud. Thus $A_{\rm V, 1D}$ measures the amount of radiation reaching the grain (see Section 3.2 for more).
Figure~\ref{ld_aalg} shows $a_{{\rm alg}}$ as a function of $A_{\rm V, 1D}$ for four densities $n_{\rm H}$.

Cho and Lazarian (\cite{cholazarian}) note that their goal is only to obtain a crude estimate. They make a couple of simplifications while deriving the above formula. First of all, they consider the anisotropy of the radiation only along the magnetic field, and neglect the isotropic radiation. They use the anisotropy factor $\gamma = 0.7$ for their dark cloud, in accordance with Draine and Weingartner (\cite{drwe96}). Our model cloud, as later shown, is much different in structure, and has large volumes of optically thin medium which are illuminated more or less isotropically. By using Eq.~\ref{eqaalg} in those regions, we overestimate the grain alignment efficiency. However, as our main goal is to examine the cores within the larger cloud, for which the anisotropy factor would be valid, we decide to use Eq.~\ref{eqaalg}. We will return to the topic of anisotropy in Section 5.1. The relative attenuation of radiation field at different wavelengths, and thus Eq.~\ref{eqaalg}, depends on the grain size distribution. The study of those effects is beyond the scope of this paper, and Eq.~\ref{eqaalg} should be considered, as said before, just an approximation.

The polarization reduction factor is
\begin{eqnarray}
R = \frac{\int_{a_{{\rm alg}}}^{a_{{\rm max}}} \; C_{{\rm ran}}n(a) \; da}{\int_{a_{{\rm min}}}^{a_{{\rm max}}} \; C_{{\rm ran}}n(a) \; da},
\label{eqR}
\end{eqnarray}
where $C_{{\rm ran}}$ is the polarization cross section of a randomly aligned grain, $n(a)$ the grain number density, $a$ the
grain size, $a_{{\rm min}}$ the minimum size of the grains, $a_{{\rm max}}$ the maximum
size, and $a_{{\rm alg}}$ the minimum aligned grain size.

\subsection{Polarized Thermal Dust Emission}

The polarized thermal dust emission is calculated 
following the formalism in Fiege \& Pudritz (\cite{fipu00}). Self-absorption
and scattering can be neglected at submillimeter wavelengths.

The Stokes Q and U components are equal to the integrals

\begin{eqnarray}
q & = & \int \alpha f \cos 2\psi \cos^2 \gamma \; ds \;,\label{eqq}\\
u & = & \int \alpha f \sin 2\psi \cos^2 \gamma \; ds \;,
\label{equ}
\end{eqnarray}
where $\alpha$ is a coefficient of the particle properties to be defined
later, $f$ a weighting function related to the intensity of the local dust emission, $\psi$ the angle between the projection of $\bf{B}$ on the plane of the sky and the north, and $\gamma$ the angle between the local $\bf{B}$ vector and the plane of the sky. In Padoan et al. \cite{paal01} the weighting function $f$ was the local gas density $\rho$, while here we consider also the case where the weighting function is the true dust emission intensity at the given wavelength, obtained from the radiative transfer calculations.

The polarization angle $\chi$ is given by
\begin{eqnarray}
\tan 2\chi= \frac{u}{q}.
\end{eqnarray}
and the degree of polarization $P$ is 
\begin{eqnarray}
P = \frac {\sqrt{q^2 + u^2}}{\Sigma - \Sigma_2} \;,
\label{eqP}
\end{eqnarray}
with
\begin{eqnarray}
\Sigma = \int f \; ds \;,
\end{eqnarray}
and 
\begin{eqnarray}
\Sigma_2 = \frac{1}{2} \int \alpha f (\cos^2 \gamma - \frac{2}{3}) \; ds \;.
\end{eqnarray}
The coefficient $\alpha$ is defined as
\begin{eqnarray}
\alpha = R F \frac{C_{\rm pol}}{C_{\rm ran}},
\label{eqalpha}
\end{eqnarray}
where $R$ is the Rayleigh polarization reduction factor due to imperfect grain
alignment, $F$ is the polarization factor due to the turbulent component of
the magnetic field, $C_{\rm pol}$ is the grain polarization cross section and
$C_{\rm ran}$ is the average cross section of a randomly aligned grain. For
more details, see Lee \& Draine (\cite{leedraine}). 

In our study, $F = 1$ because the three-dimensional magnetic field is given by
the MHD models and we assume that the small-scale structure is resolved in the
numerical solution. The ratio $C_{\rm pol}/C_{\rm ran}$ needs to be fixed,
because we want to study the effects caused by variations of $R$.  In order to
avoid unreasonably high polarization degree, we choose $C_{\rm pol}/C_{\rm
ran} = 0.15$. This corresponds to the axial ratio of roughly 1.1 (see Fig. 1
in Padoan et al. \cite{paal01}). In our study, we use both uniform $R=1$ as
well as $R$ which varies according to the local efficiency of the radiative
torque, as explained in Section 2.1.

\section{Models}

In this section, we describe the MHD models and the radiative transfer
calculations that are used in the simulation of the dust polarization.

\subsection{MHD models}

In this work, we use the three numerical models - $A$, $B$, and $C$ - which
are described in detail in Juvela, Padoan \& Nordlund (\cite{jual01}). They
are based on the results of numerical simulations of highly supersonic
magnetohydrodynamic turbulence, run on a $128^3$ computational mesh with
periodic boundary conditions. The initial density and magnetic field were
uniform and an external random force was applied to drive the turbulence at
roughly constant rms Mach number of the flow. All three models have highly
supersonic flows with the rms sonic Mach number $M_{\rm s} \sim 10$. Models
$B$ and $C$ are super-Alfv\'enic, with $M_{\rm A} \sim 10$, while model $A$
has equipartition of magnetic and turbulent kinetic energy, $M_{\rm A} \sim
1$, and thus an order of magnitude stronger magnetic field than the other
two. Model $C$ includes also the effect of self-gravity. Details of
the numerical code used to solve the MHD equations can be found in Padoan \&
Nordlund (\cite{pano99}).

\subsection{Radiative transfer calculations}

In our scheme radiative transfer calculations are needed for two purposes.
First, the radiative transfer is solved in order to compute the dust 
temperature and the resulting emission, based on a Monte Carlo code 
discussed in detail elsewhere (Juvela \& Padoan \cite{Juvela2003}; 
Juvela \cite{Juvela2005}). The total intensity of the dust emission is used 
later to analyze the $P/I$-relations. Once the total intensity
and polarization are computed, radiative transfer effects can be ignored, 
because the dust emission is optically thin at the wavelengths considered 
($\lambda>100\,\mu$m). Second, the radiative transfer is also calculated
in order to estimate the three-dimensional distribution of the radiation
field inside the model clouds, which is needed to compute the minimum
grain size for alignment. This simulation is done in the V-band
(0.55$\mu$m) and the computed ratios between the local and the background
intensities, $I/I_{\rm bg}$, are transformed into $A_{\rm V, eff}$
according to the formula
\begin{equation}
I/I_{\rm bg} = exp(-1.086 \times A_{\rm V, eff}).
\label{eqaeff}
\end{equation}
The values of $A_{\rm V, eff}$ are then used to estimate the minimum 
size of aligned grains according to Eq.~\ref{eqaalg}. However, this $A_{\rm V, eff}$ is not the same $A_{\rm V, 1D}$ as used by Cho \& Lazarian (\cite{cholazarian}). $A_{\rm V, eff}$ is a measure of the amount of external radiation reaching a grain (see Eq.~\ref{eqaeff}). In the centre of a homogeneous and spherically symmetric cloud $A_{\rm V, eff}$ would be equal to $A_{\rm V, 1D}$ (half of the $A_{\rm V}$ through the whole cloud). In inhomogeneous clouds the radiation propagates preferentially along low $A_{\rm V}$ sightlines. Therefore, the  value $A_{\rm V, eff}$ can be significantly smaller than the average $A_{\rm V}$ that would be calculated as an arithmetic average over all directions. As $A_{\rm V, 1D}$ is calculated along the lowest $A_{\rm V}$ sightline (i.e. perpendicular to the surface), the actual amount of radiation reaching the cell is less than in the case where $A_{\rm V, eff}$ would have the same numerical value. Thus, a given value of $A_{\rm V, eff}$ needs to be transformed into a lower value of $A_{\rm V, 1D}$ in order to derive the same radiation field. This is done in our case by numerically integrating over all directions to derive the actual total radiation reaching the cell as a function of $A_{\rm V, 1D}$. This allows the matching of the corresponding values of $A_{\rm V, 1D}$ and $A_{\rm V, eff}$.

In the calculations, the dust properties correspond to normal Milky Way 
dust with $R_{\rm V}=$3.1 (Weingartner \& Draine \cite{wedr01}) and we use the
scattering functions calculated by Draine (\cite{Draine2003})\footnote{The
dust optical properties are available on the web at
www.astro.princeton.edu/$\sim$draine}. The cloud models are scaled to a
linear size $L = 6 \; {\rm pc}$ and a mean density $n = 640 \; {\rm cm^{-3}}$. 
The radiative transfer calculations depend only on column density, 
which is $<N(H)>=1.2\times 10^{22}$cm$^{-2}$, corresponding to $<A_{\rm
V}> \sim 6^{\rm m}$. However, the absolute gas density enters the
calculations through the density dependent polarization reduction factor $R$
(see next Section).

\section{Results}

\subsection{Polarization maps} \label{sect:pol_maps}

\begin{figure*}
\centering
\includegraphics[width=16cm]{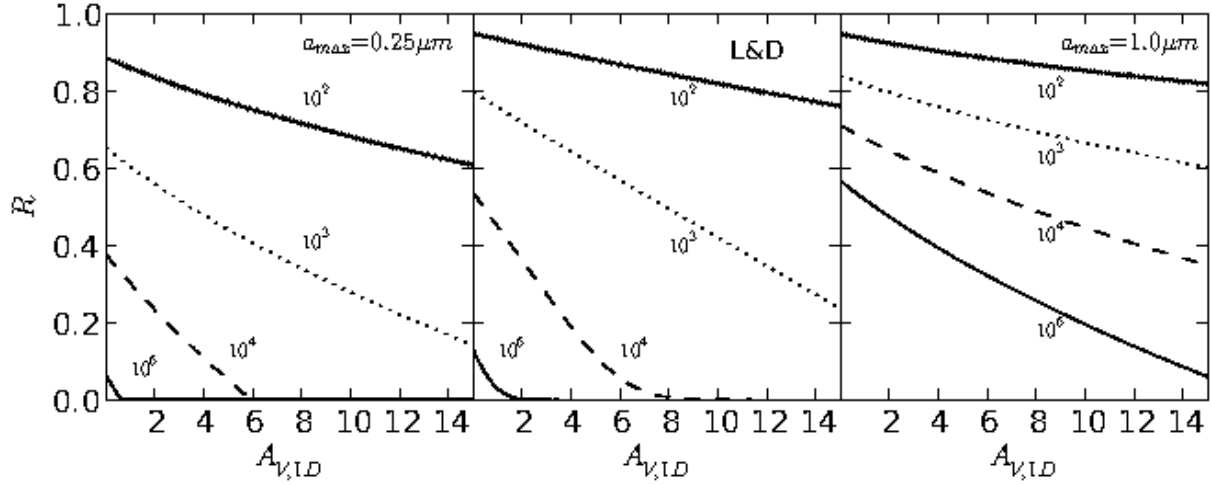}
\caption{
Polarization reduction factor $R$ as a function of $A_{\rm V, 1D}$ using Weingartner \&
Draine (\cite{wedr01}) dust model. The curves correspond to different
densities. The frames correspond to different values of the maximum grain
size, $a_{\rm max}$. In the first and the third frame the grain size
distributions are pure power laws, $n(a)\sim a^{-3.5}$, while in the middle
frame the size distribution is taken from Li \& Draine (\cite{Li2001}).
}
\label{Rfactors}
\end{figure*}

We calculate $R$ in the following manner. First, we transform $A_{\rm V, eff}$ to the corresponding $A_{\rm V, 1D}$ and use $A_{\rm V, 1D}$ to calculate $a_{\rm alg}$ from Eq.~\ref{eqaalg} for each cell. We set $a_{\rm min} = 0.005
\;\mu{\rm m}$. We use two power-law distributions, $n(a) \propto a^{-3.5}$,
with sharp cut-offs at $a_{\rm max} = 0.25 \;\mu{\rm m}$ and $a_{\rm max} =
1.0 \;\mu{\rm m}$. As a third grain size distribution we also use the model of
Li and Draine (\cite{Li2001}), hereafter L\&D, which is not a pure
power law and has an effective cut-off between the previous
$a_{\rm max}$ limits. Eq.~\ref{eqR} is solved numerically to obtain the values of the resulting polarization 
reduction factors for a grid of values of $n_{\rm H}$ and $A_{\rm V, 1D}$. 
In the actual cloud model, the $R$-values are obtained for each cell by
interpolating from the pre-calculated grid, rather than by solving
Eq.~\ref{eqR} for each cell separately. The polarization reduction factor $R$
as a function of $A_{\rm V, 1D}$ are shown in Fig.~\ref{Rfactors} for the
different grain size distributions and different densities.  As the maximum
grain size increases, the $R$ factor becomes less dependent on both the
density and the value of the  extinction, $A_{\rm V, 1D}$. This is the
direct consequence of Eq.~\ref{eqR}, as $a_{\rm alg}$ reaches $a_{\rm max}$
less rapidly.

In the calculations of the polarization maps, we use model $C$ as our
baseline, which is scaled to the size $L = 6 \; {\rm pc}$ and mean density $n
= 640 \; {\rm cm^{-3}}$. This results in an optically thick cloud with average
visual extinction $<A_{\rm V}>\sim 6.3^{\rm m}$ through the cloud. We also choose 353\,GHz as
our base frequency. Other frequencies and model clouds are examined later.

\begin{figure*}
\centering
\includegraphics[width=16cm]{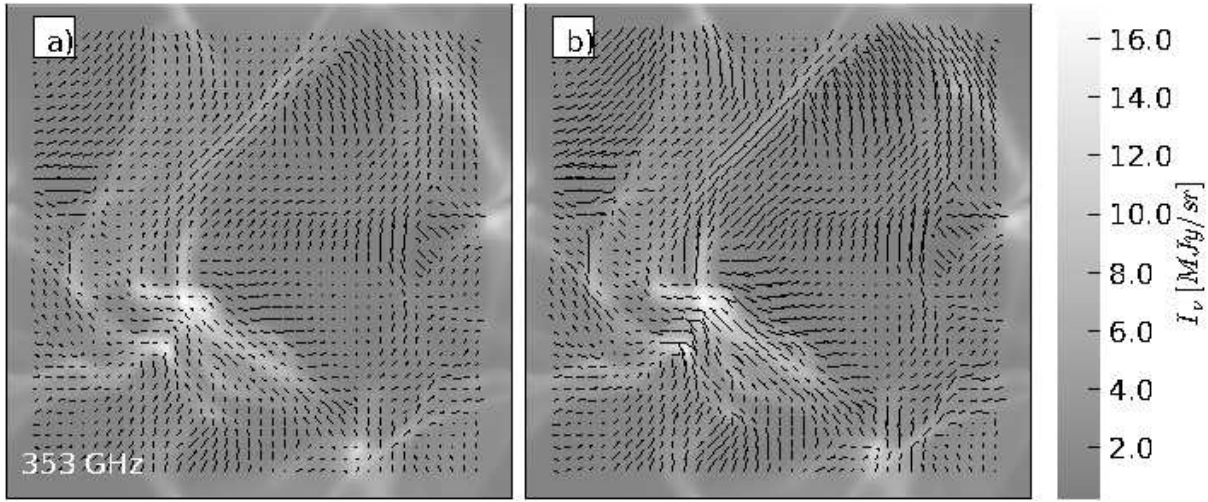}
\caption{
Simulated polarization maps at 353\,GHz with variable $R$ (left frame) and
uniform $R=1$ (right frame). Polarization vectors are drawn for every third
pixel, and the scaling of the vector lengths is the same in both frames. In
the left frame the maximum polarization degree is 9\% while in the right frame
it is 13\%. The background image shows the total intensity at this frequency.
}
\label{ld_emitted_353}
\end{figure*}

Figure~\ref{ld_emitted_353} shows the calculated polarization maps for
353\,GHz. The polarization is calculated using the actual three-dimensional
distribution of dust emission, that is using the local dust emission at the
given frequency as the weighting function $f$ (see Eq.~\ref{eqq}).  In the
left frame the polarization reduction factor $R$ varies according to the local
values of density and $A_{\rm V, 1D}$, as shown in the middle frame of 
Fig.~\ref{Rfactors}. In the right hand frame a uniform value of $R=1$ was
assumed. The difference in the polarization degrees is clearly visible. In
particular, with variable $R$ the polarization degree decreases in regions of
high intensity, while no such trend exists in the right hand frame.

\begin{figure}
\centering
\includegraphics[width=8cm]{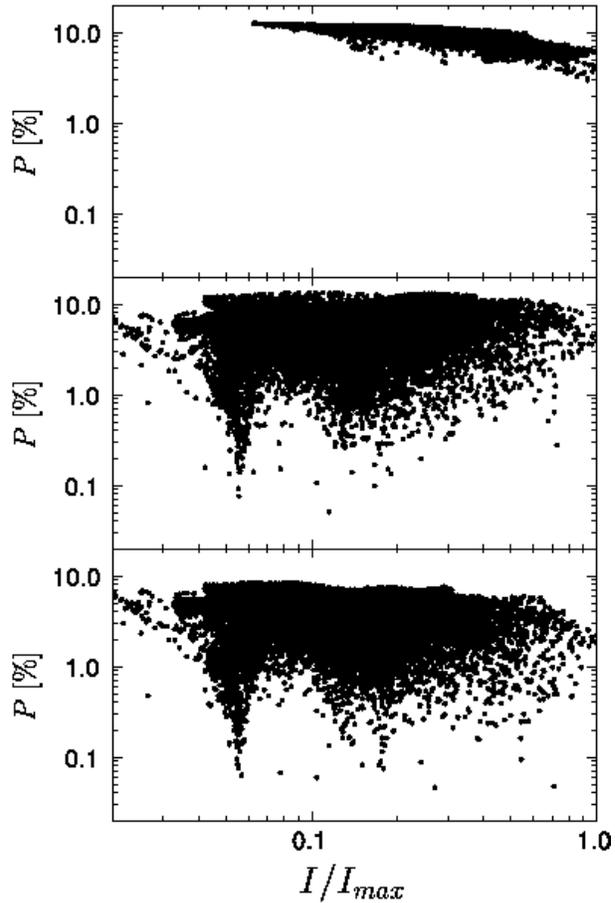}
\caption{
Polarization degree for the whole $128^2$ pixel maps, viewed from the
$y$-direction. The uppermost frame corresponds to model $A$ where the variable
$R$ factors are calculated using Li \& Draine (\cite{Li2001}) grain size distribution. The other frames
are for model $C$ with uniform $R=1$ (middle frame) and variable $R$ (lower
frame), respectively.
}
\label{ld_poldeg_353}
\end{figure}

In Fig.~\ref{ld_poldeg_353} the polarization degree and intensities are shown
for the whole maps. The shapes of the plots are quite similar for the two
lowermost frames, which are for model $C$ with constant and variable $R$
factors.  The wide range of polarization degrees at a any given intensity is
caused by the geometry of the magnetic field and not by any numerical noise.
This can be clearly seen when comparing the results with the uppermost frame,
which is for model $A$. Model $A$ has a strong, structured magnetic field in
the z-direction, with only small deviations. Thus, the depolarization effects of
tangled magnetic fields in the line of sight are small, leading to a much
tighter relation between $P$ and $I$. 

On closer look the two lowermost frames in Fig.~\ref{ld_poldeg_353} show some
clear differences. First, there is a reduction of the maximum polarization
degree from 13\% to 9\%. This is easily explained if one looks at the middle
frame of Fig.~\ref{Rfactors} and recalls that we are considering a quite dense
cloud, leading to $R < 1$ even in the low-density parts of the cloud. The 
low-intensity points to the extreme left are unaffected, but otherwise the
polarization degree is clearly reduced, even for low intensity sightlines.

Another noticeable feature is the change in the upper envelope of the
polarization degree. This can be seen by comparing the lower frames.  With
variable $R$, on sightlines with higher intensity (and higher average volume
density) the maximum polarization degree becomes gradually smaller.  In
model $C$ the effect is clear, but still small compared with the overall
scatter in the polarization degrees. The overall slope is seen more clearly in
in the case of model $A$ (the uppermost frame), because of the smaller
scatter. The maximum polarization drops from about 12\% to 6\% in the interval
$0.06 < I/I_{\rm max} < 1.0$. These effects will will be
discussed further in the next subsection.

\subsection{Self-gravitating cores}

\begin{figure*}
\centering
\includegraphics[width=16cm]{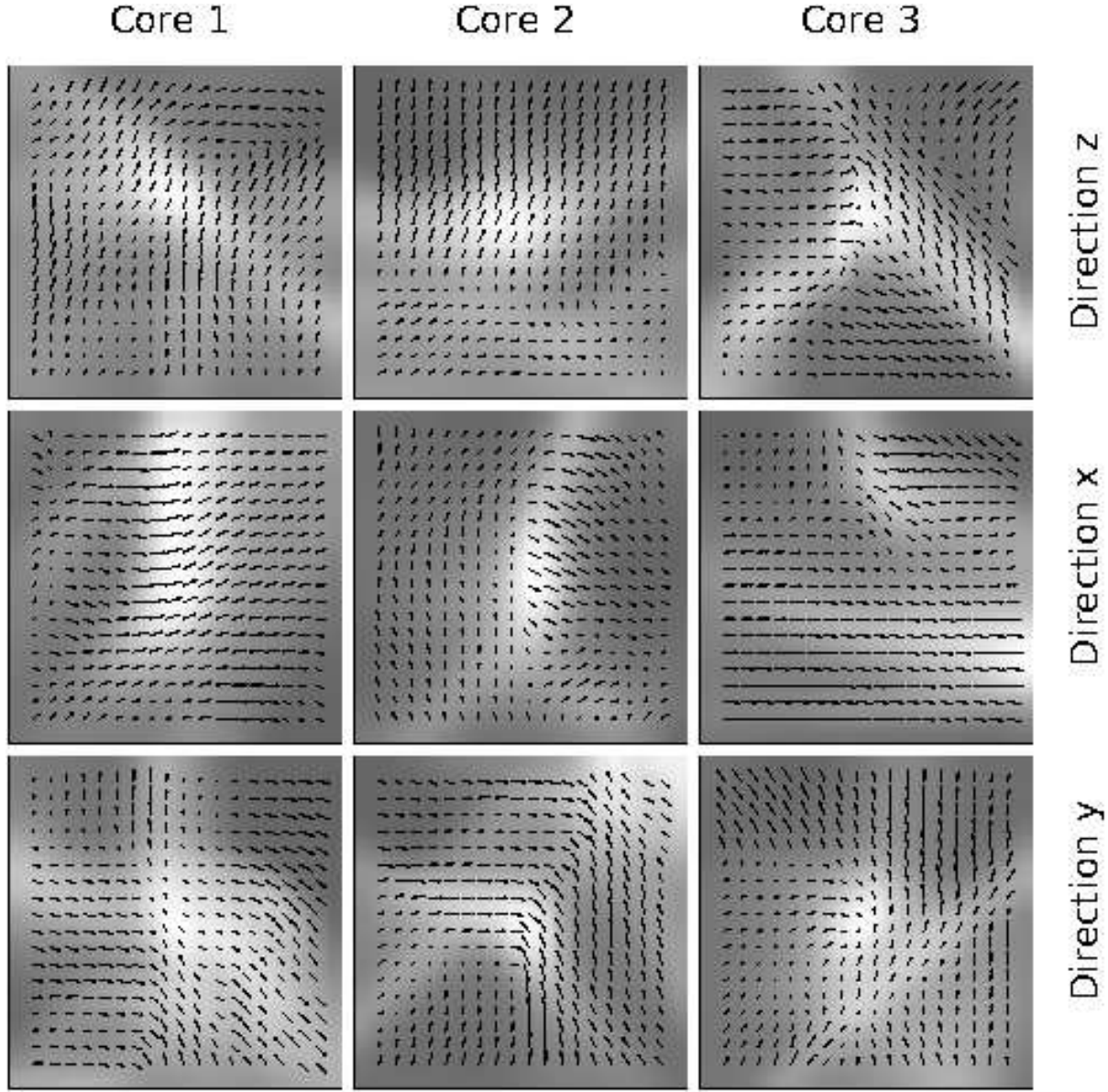}
\caption{
Selected cores of model $C$ observed at 353\,GHz. The background grayscale map
shows the intensity, and the vectors the direction and degree of polarization
for the case of $R = 1$.
}
\label{noRcores}
\end{figure*}

\begin{figure*}
\centering \includegraphics[width=16cm]{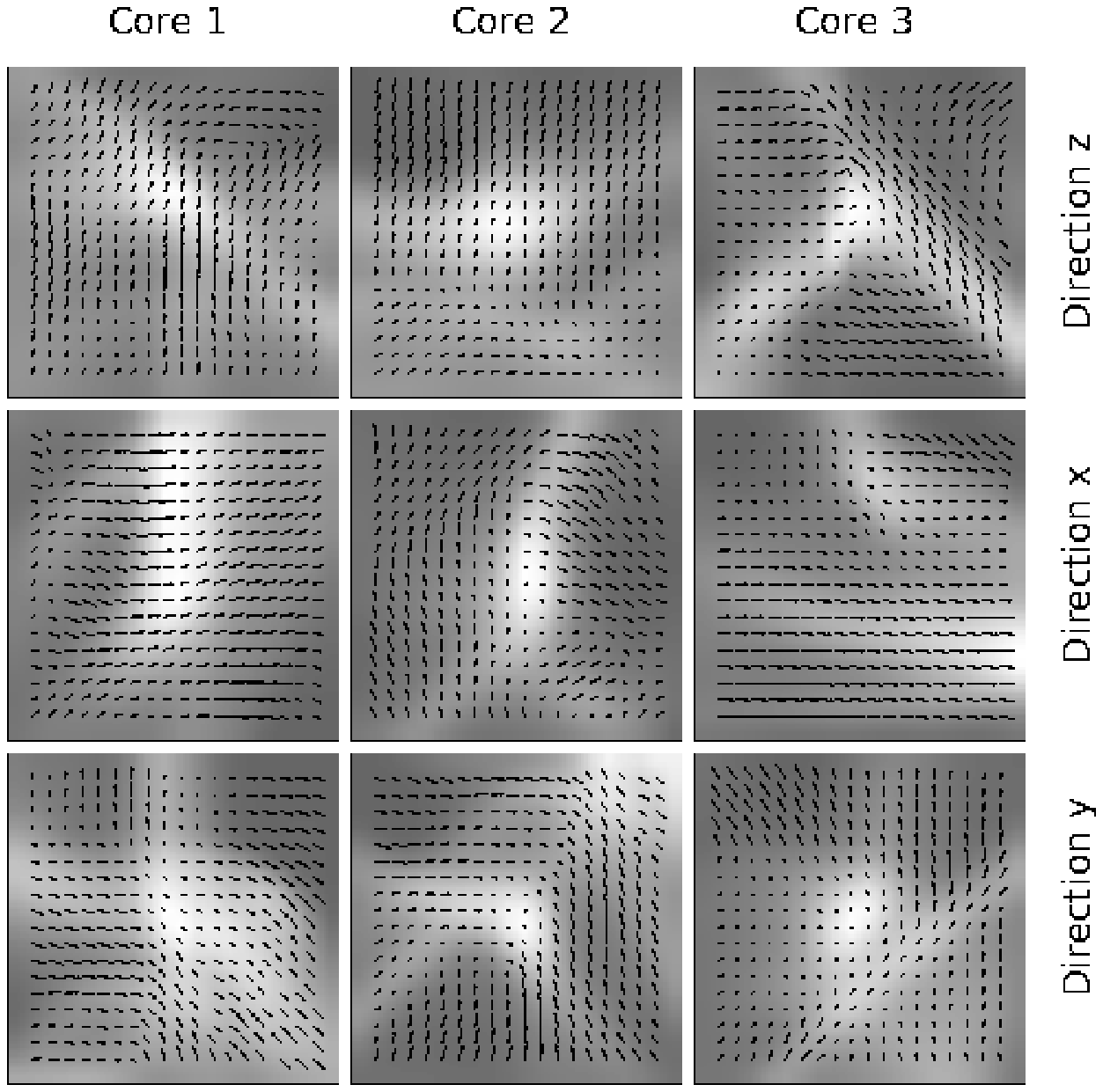} 
\caption{
Selected cores of model $C$ at 353\,GHz. The figure is identical with
Fig.~\ref{noRcores}, but assumes variable polarization reduction factor, $R$.
}
\label{cores}
\end{figure*}

Figures~\ref{noRcores} and ~\ref{cores} show zoomed 20 pixel by 20 pixel maps
of selected cores of model $C$ clumps. Maps are shown for both $R = 1$ and
variable $R$. The model $C$ included the effect of self-gravity. However, what
is interpreted as a core in observations need not be a clear-cut, compact
gravity bound cores in reality. The same applies to these objects that were
selected based on observations in the direction $y$. These are knots in
the filamentary cloud structure, and in certain directions the structures can
be very elongated.

The general trend towards smaller polarization degree in the cores is obvious.
The relation between observed total intensity and polarization degree is,
however, not always clear. Density enhancements close to cloud edges provide
one exception. Because they are subjected to unattenuated external radiation
field, they can exhibit large emission although they have relatively low
density compared with actual cores. Thus, the factor $R$ is larger than in
the cores and the polarization very similar to the case of uniform $R$. One
example of this can be seen in Core 3, when Figs.~\ref{noRcores} and
~\ref{cores} are compared. This is particularly striking in the direction $x$,
where the actual core is situated in an area of lower brightness slightly
above the map centre, while a bright filament on the cloud surface crosses the
lower part of the map. Another thing to note is that the reduction of the
polarization in the dense parts of the cloud can, in special cases, actually
increase the total polarization degree. The core 2 provides an example of this, when
observed from the $x$-direction. Although the polarization decreases in the
bright core itself, the polarization increases immediately to the left of it.
This is, of course, due to the magnetic field geometry. If the magnetic field
is predominantly of opposite direction in high and low density regions on the
line of sight, they depolarize the observed intensity. When the polarization
reduction factor reduces the polarization in the dense region, the
depolarization effect is reduced and the observed net polarization is
correspondingly increased. This is another indication of the fact that it is
difficult to interpret observations of polarization if not in a
statistical sense.

\begin{figure*}
\centering
\includegraphics[width=16cm]{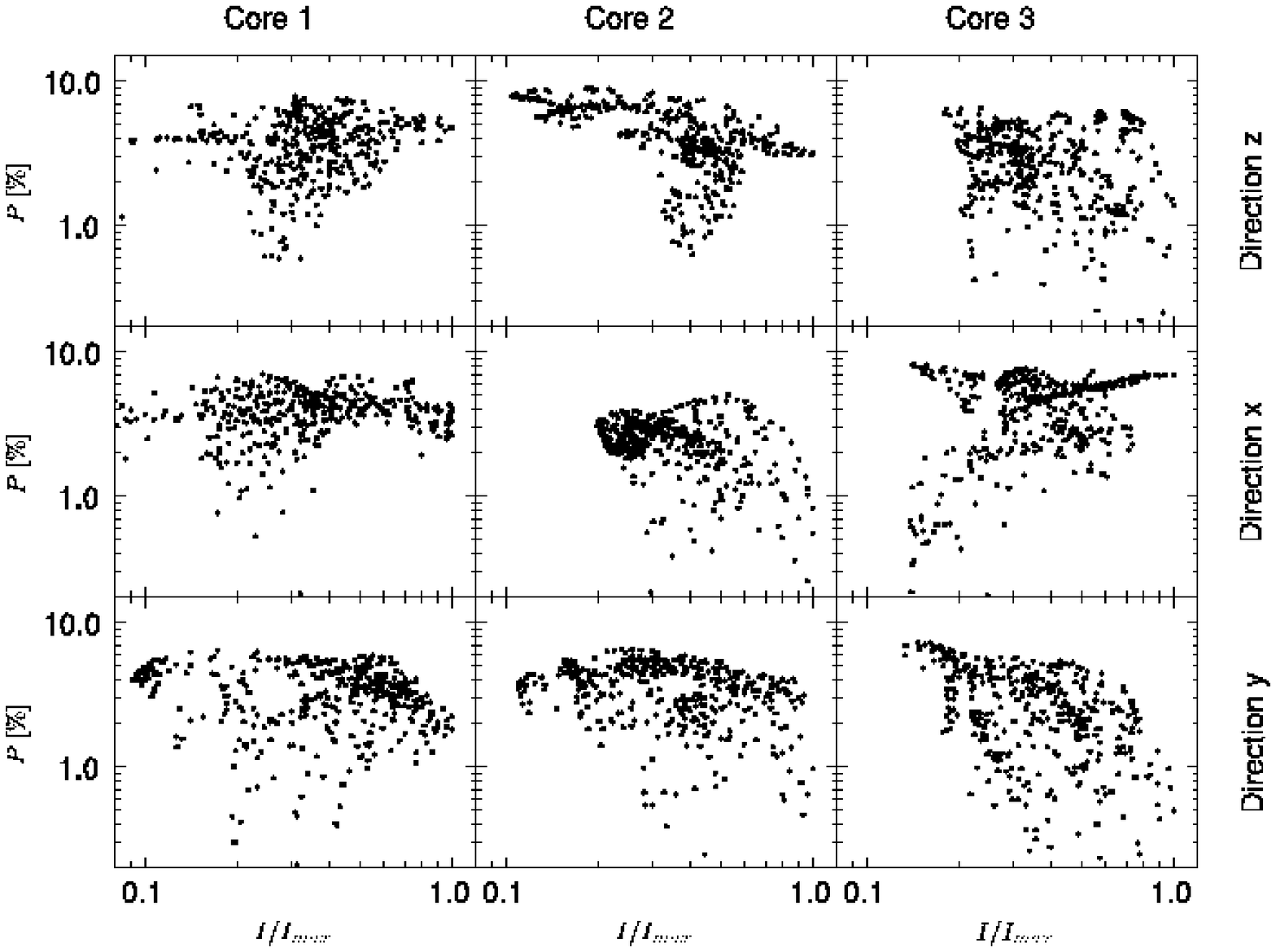}
\caption{
The relation between polarization degree and total intensity for selected
cores of the model $C$. The polarization degree has been calculated using the
WD2001 dust model and variable polarization reduction factor, $R$.
}
\label{corepol}
\end{figure*}

Fig.~\ref{corepol} shows the plotted $P/I$-relations for the cores seen in
Fig.~\ref{cores}. In the direction $x$ plot of Core 3, one can clearly
identify the bright, high polarization filament on the edge of our cloud.
Other bright but less dense (thus high polarization) regions help to mask the
slope. The observed polarization is a sum of many effects, depolarization
being the most obvious one. For example, the large-scale structure of the
cloud determines which parts of the cloud receive the most radiation from the
external radiation field. This affects the efficiency of the radiative torque
and of grain alignment (taken into account in $A_{\rm V, 1D}$ in
Eq.~\ref{eqaalg}) and also heats up the dust grains, which increases their 
thermal emission. In the plots, the scaling of the $x$-axis depends on the
maximum intensity which, therefore, affects the appearance of the figures. One
example of this is Core 2 seen in the direction $y$, where one can easily see the
bright upper right corner of Fig.~\ref{cores} represented as the cluster of
high polarization, high intensity points in Fig.~\ref{corepol}. If this area
were omitted, the drop of polarization degree with intensity would become
much more dramatic.

\begin{figure}
\centering
\includegraphics[width=8cm]{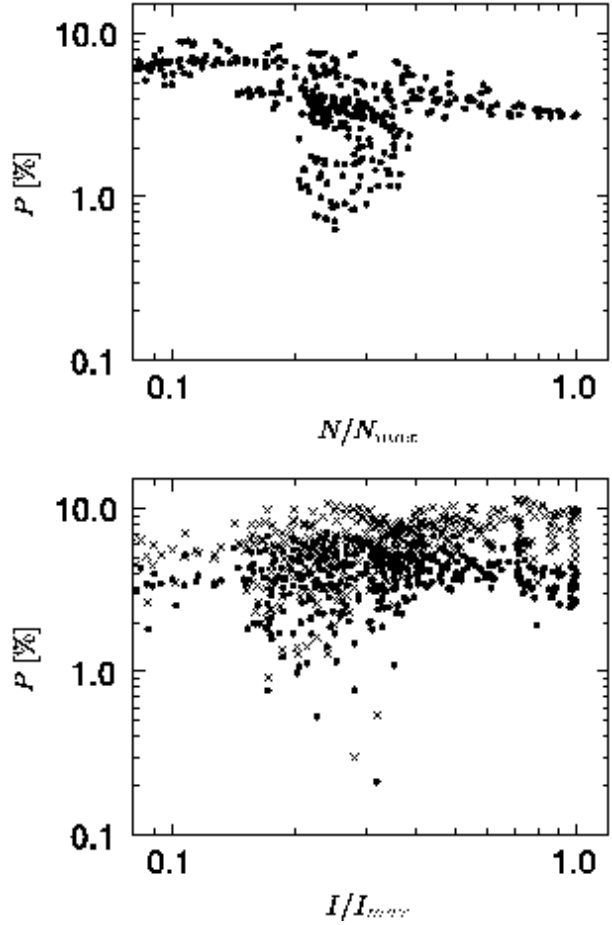}
\caption{
Selected cases for comparison with Figs.~\ref{noRcores} and \ref{cores}. The upper
frame shows the polarization degree for Core 2 as a function of true column
density, when viewed from the direction $z$. The lower frame shows Core 1 viewed
from the direction $x$, with variable $R$ (dots) and uniform $R=1$ (crosses).
The polarization degree is plotted as a function of the total intensity at the
same frequency of 353\,GHz.
}
\label{corepoldeg}
\end{figure}

In Fig.~\ref{corepoldeg} we look at some of these effects. In the uppermost
frame we show for the Core 2 the polarization degree in the direction $z$ as
a function of column density $N$. The clump of low polarization points,
identified with the filament in the lower portion of the image in
Fig.~\ref{cores}, moves from moderate intensity to low column density. Thus
these points are not part of any high-density core, and the low polarization
is likely to be caused by the depolarization resulting from the magnetic field
structure. If those points were omitted, we would end up with a quite clean
$P/I$-relation for the core itself. Of course, the problem with this kind of
approach is that, for an observer, the true column density is not a priori
known. However, if accurate column density estimates were made, this
kind of point-to-point investigation would help in distinguishing between true
cores and smaller density enhancements on the cloud surface. If the magnetic
field is preferentially parallel to the filament, the result, high intensity and
low polarization, mimics the appearance of a dense core with small value of
$R$. There are some example of the opposite effect in the cores, where the
fainter core points move in the plots to the right when we shift from the
intensity to the column density axis. Long filaments that are aligned along the
line-of-sight would produce a relatively high surface brightness compared with
the actual column density.

In the lowermost frame of Fig.~\ref{corepoldeg} the difference between the
uniform $R=1$ case and the variable $R$ depending on density and $A_{\rm V, 1D}$ is obvious. With $R=1$ the maximum polarization increases, rather
than decreases, to $\sim$10\% with increasining intensity. When $R$ is
variable there is a downward trend, ending at around 3\% at the brightest
regions. This is explained by the fact that intensity and density are usually
connected. The high-intensity points correspond to the high column density. In
the high density regions we also expect to find high $A_{\rm V, 1D}$.
Thus, looking at the middle frame of Fig.\ref{Rfactors}, we see that the
polarization reduction factor $R$ decreases quite rapidly as we go to denser
and more opaque regions. Therefore, those regions affect the polarization
less, whereas in the $R=1$ case, it is precisely the dense parts that
contribute the most.

\subsection{Effects of Different Cloud and Dust Models}

In this section, we examine differences between different cloud models as well
as discuss the possible effects of grain growth.
We will also briefly consider the effect of the choice of wavelength in our
polarization calculation.

\begin{figure*}
\centering
\includegraphics[width=16cm]{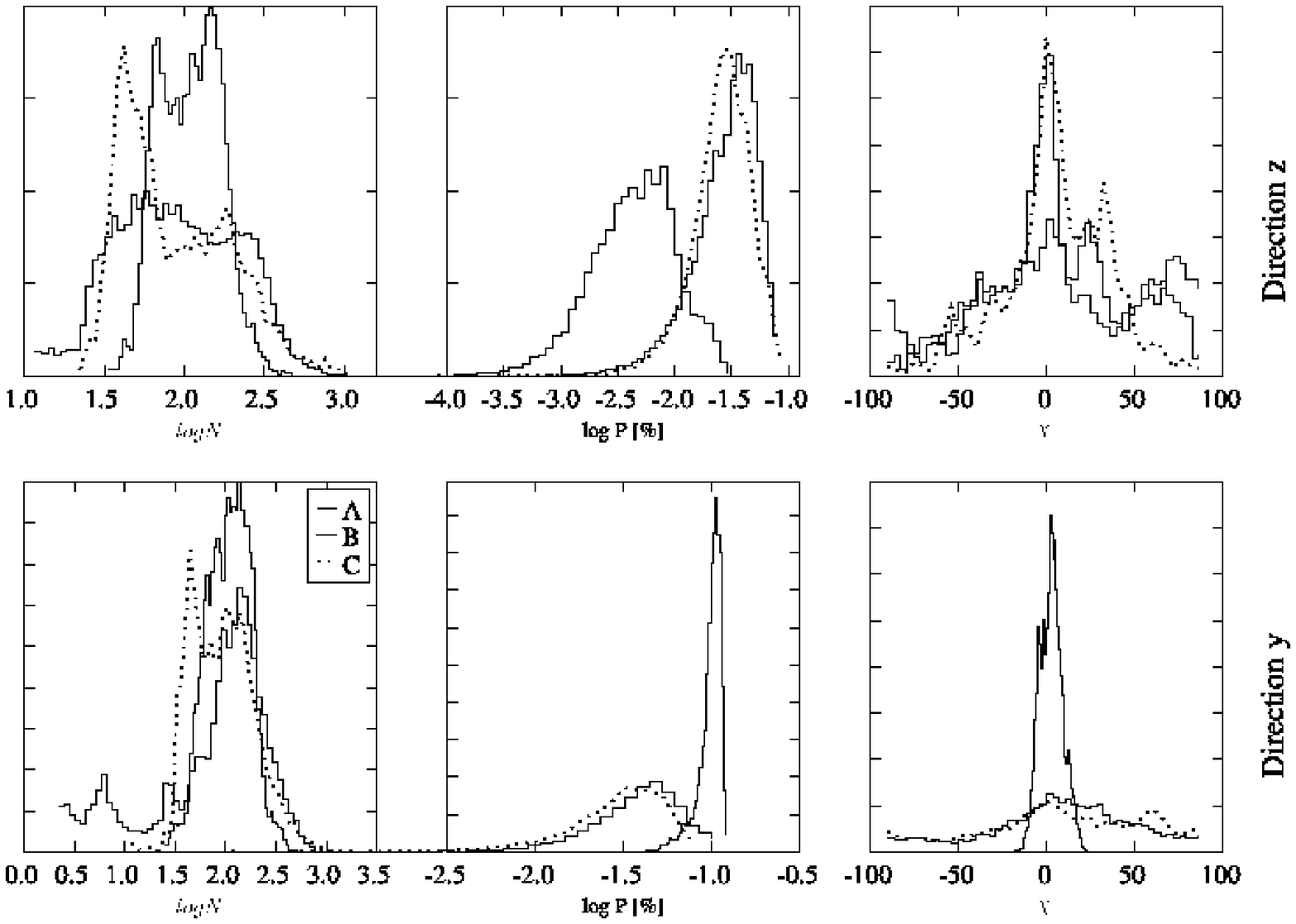}
\caption{
Histograms of column density, polarization degree, and polarization angle for
the three model clouds $A$, $B$, and $C$.  The histograms of the polarization
angle $\chi$ have been computed so that the maximum value is at zero angle.
}
\label{models}
\end{figure*}

Figure.~\ref{models} shows some histograms for out three model clouds.
Particularly noticeable is the sharp contrast when the polarization degree and
angle of model $A$ are compared with the other models $B$ and $C$.
Similarly, in model $A$ there is a very strong dependence on the direction of
observations. These results from the fact that model $A$ has a very strong
magnetic field that is aligned along the $z$-direction. There is little
depolarization caused by magnetic field tangling, and, consequently, the
polarization degree is high, and the angle of the polarization vectors is
almost uniform. When model $A$ is observed from the $z$-direction, the angle
$\gamma$ is close to perpendicular to the plane of the sky and hence $\cos^2 \gamma$ is small (see Section 2.2). Thus, the average polarized signal is small, and small variation in field geometry can lead to strong depolarization. However, because there is an obvious preferred angle even in $z$-direction, we can say that the evolved magnetic field is not aligned perfectly along the $z$-axis, but has a small deviation from it.

Models $B$ and $C$ are both super-Alfv\'enic and, consequently, the field
geometry is much more complicated. Comparing these two models, we notice that
the polarization degree is slightly smaller in model $C$ when observed from
any direction. On the other hand, the distributions of polarization degrees
are quite similar. While model $C$ has slightly more very dense cores,
probably due to its inclusion of self-gravity, the number of dense cells is so
small that it seems implausible that $R$ would be the cause of this
difference, especially since model $B$ shows a similarly large fraction of
very dense cells. However, model $B$ seems to have a slightly less tangled
magnetic field and, therefore, less depolarization.

\begin{figure*}
\centering
\includegraphics[width=16cm]{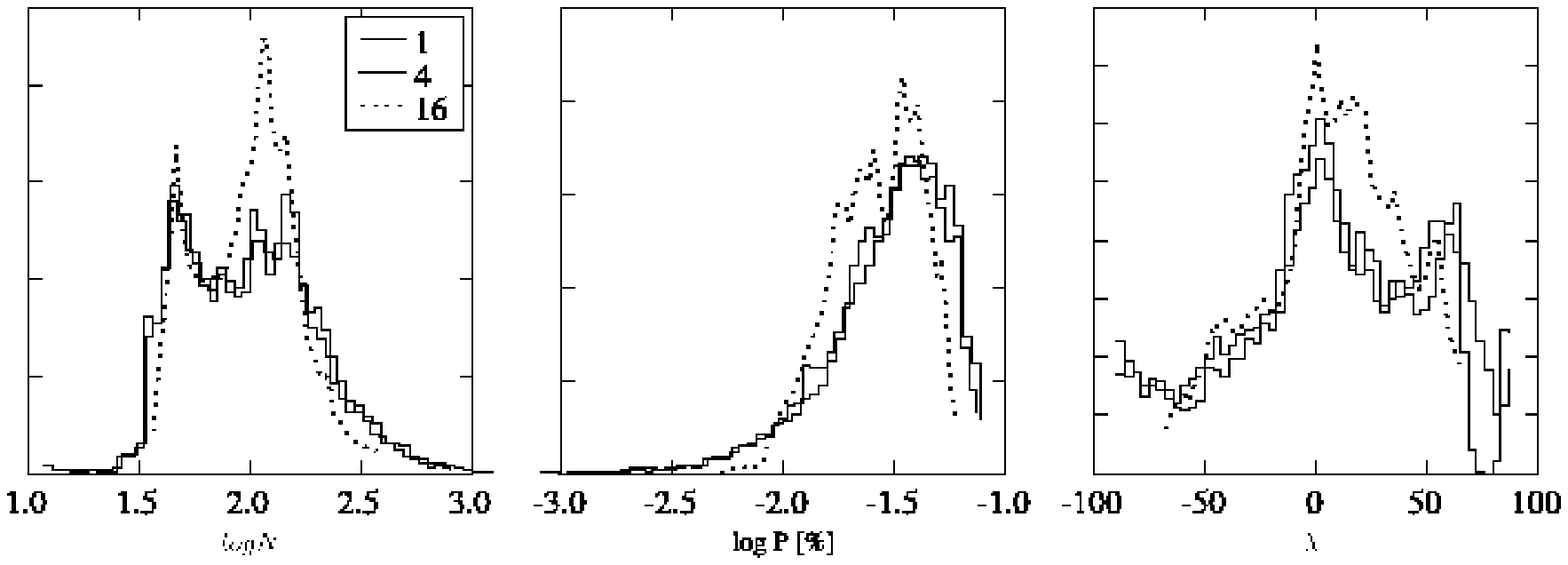}
\caption{
Histograms of column density, polarization degree, and polarization angle
computed for model $C$ viewed from $y$-direction. The three curves correspond
to different resolution of observation, corresponding to FWHM of either 1, 4,
or 16 pixels. 
}
\label{res}
\end{figure*}

In Fig.~\ref{res} we show similar histograms for model $C$ when the cloud is
observed at resolutions corresponding to 1, 4, or 16 pixels. The four-pixel
resolution seems almost identical to the original one. The 16-pixel resolution
shows clear results of the averaging, by narrowing down the variance of the
distributions. Of course, this is to be expected as soon as the beam size
exceeds the size of typical structures. If we are interested in the details,
such as the cores, the best possible resolution is, of course, needed. On the
other hand, for general morphological and statistical studies the
results are not very sensitive to a small reduction in resolution.

\begin{figure*}
\centering
\includegraphics[width=16cm]{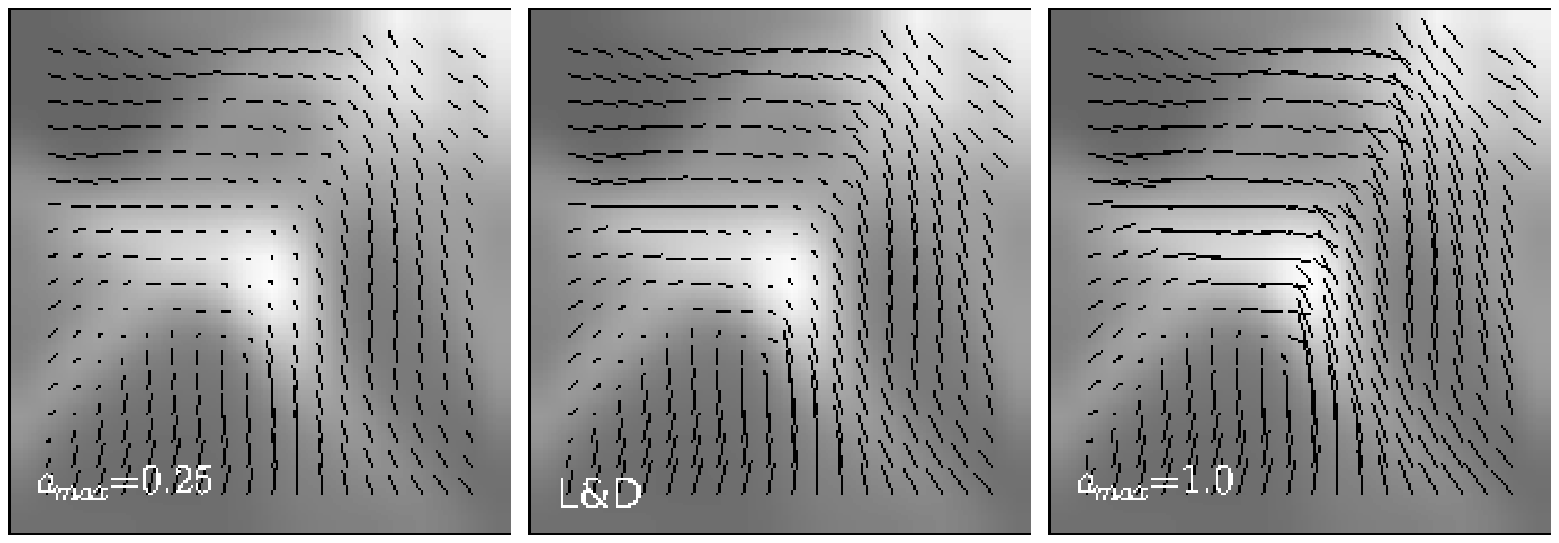}
\caption{
Polarization observed toward Core 2 in case of different values of $a_{\rm
max}$. The observations are made toward the $y$-direction. Polarization
vectors are drawn for every pixel, and the scaling of the vectors is identical
in all frames. The maximum polarization is 5.1\% in the left frame, 6.6\% in the middle and 7.8\% in the right.
}
\label{mapR}
\end{figure*}

\begin{figure*}
\centering
\includegraphics[width=16cm]{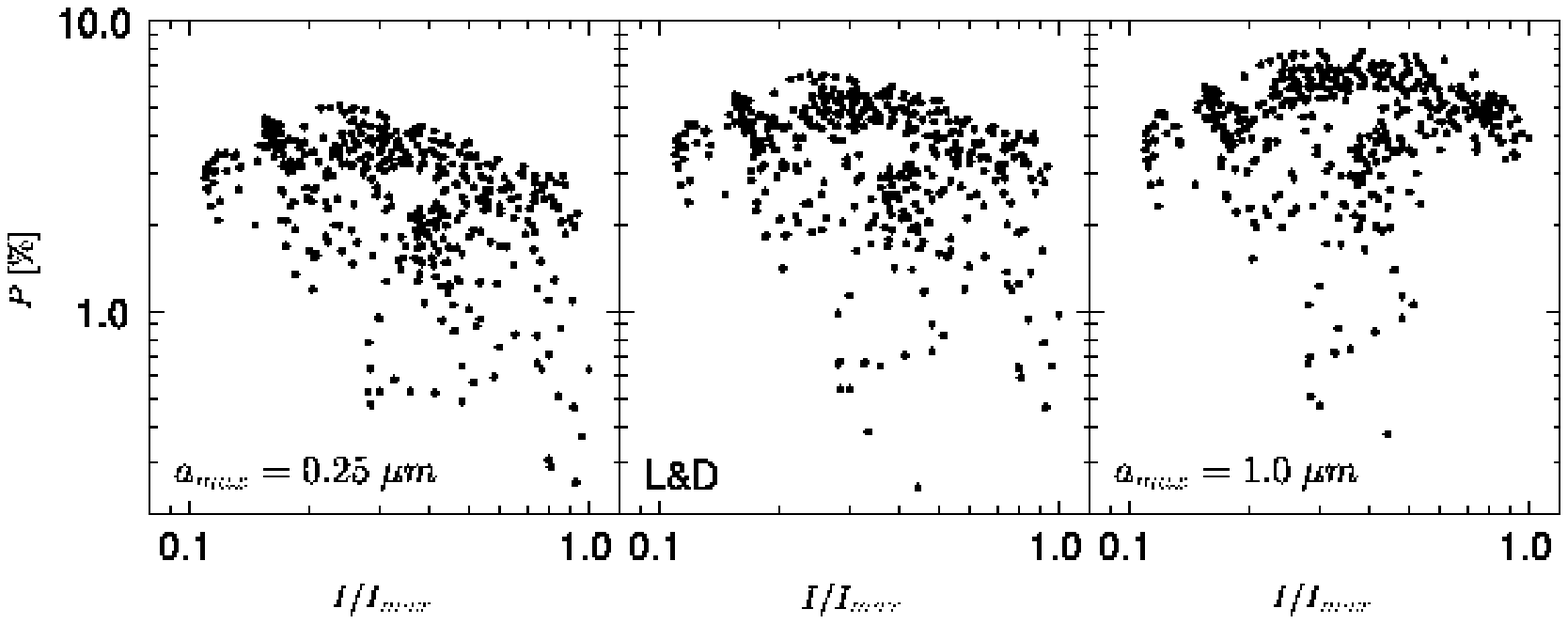}
\caption{
Polarization degree as a function of intensity for Core 2 viewed from the $y$-direction.
The frames correspond to different values of the maximum grain
size, $a_{\rm max}$. In the first and the third frame the grain size
distributions are pure power laws, $n(a)\sim a^{-3.5}$, while in the middle
frame the size distribution is taken from Li \& Draine (\cite{Li2001}).
}
\label{scatterR}
\end{figure*}

Figures~\ref{mapR} and ~\ref{scatterR} are similar to Figs.~\ref{cores} and
~\ref{corepol}, showing Core 2 in $y$-direction with the three different grain
size distributions that were introduced at the beginning of
Section~\ref{sect:pol_maps}. The increase of the polarization vectors and even
a slight change of direction is clearly visible, as we move toward larger
$a_{\rm max}$. This can be easily explained by looking at Fig~\ref{Rfactors}.
The smaller the parameter $a_{\rm max}$, the faster $R$ decreases as a
function of density and $A_{\rm V, 1D}$. This is because in
Eq.~\ref{eqR} $a_{\rm alg}$ quickly reaches the smallest $a_{\rm max}$, while
the largest $a_{\rm max}$ takes a larger range of $A_{\rm V, 1D}$ values to build
up to, hence making the transition more smooth.

In Fig.~\ref{scatterR} we see the drop in the maximum $P$, similar to the
global plots. One can also see that while the two frames on the left resemble
one another in shape, the rightmost frame is significantly different. This 
hints at the fact that at $a_{\rm max} = 1 \; {\rm \mu m}$, we are approaching
the case of $R = 1$. This indicates that the number of very dense cells is
relatively small, as can be seen from Fig.~\ref{models}, and that $A_{\rm V, 1D}$ does not reach as high values as $A_{\rm V}$, the visual extinction through the whole cloud. In an inhomogeneous cloud there are always low density
regions and 'holes' that allow external radiation field to penetrate deep into
the cloud. Therefore, $A_{\rm V, 1D}$ values can be much smaller than
the average $A_{\rm V}$, which, furthermore, is calculated through the whole cloud.
In model $C$, $A_{\rm V, 1D}$ has a maximum value of $4.3^{\rm m}$
and a mean value of $0.03^{\rm m}$. We can also see the slope of the decrease
of maximum $P$ slowly leveling off as we move to higher values of $a_{\rm
max}$.

\begin{figure*}
\centering
\includegraphics[width=16cm]{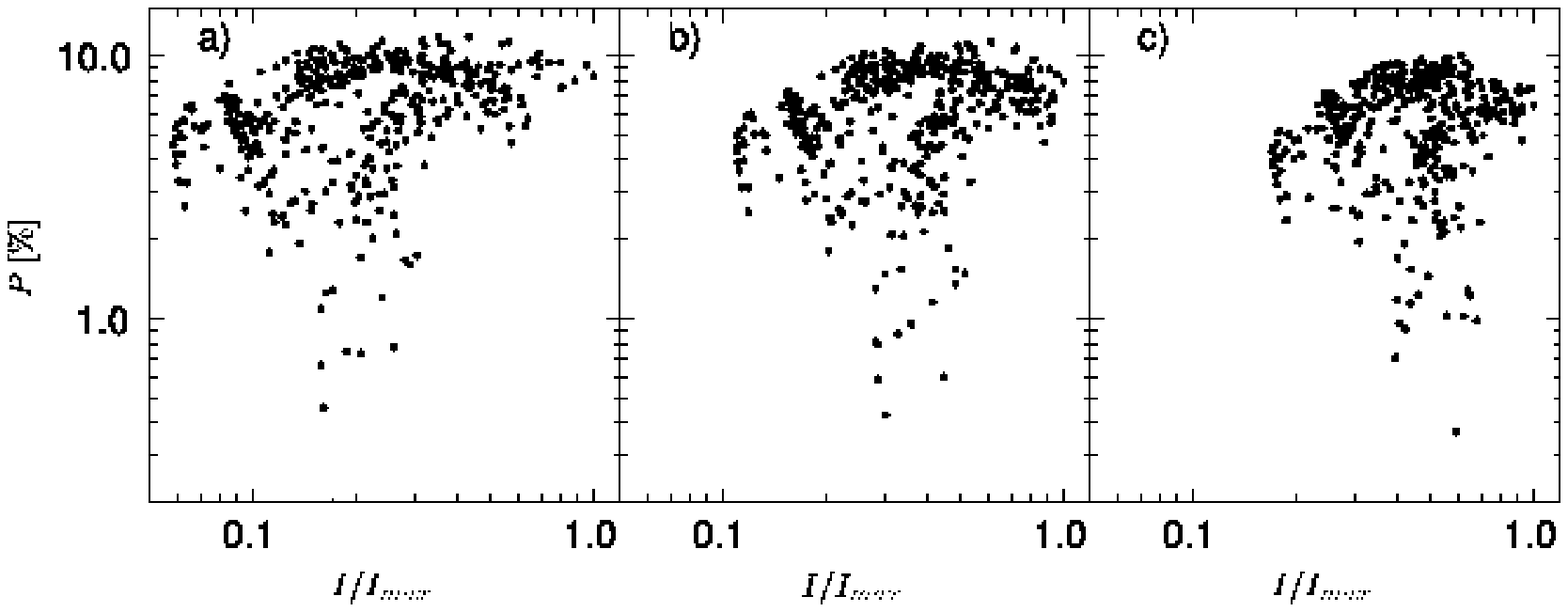}
\caption{
Polarization degree of Core 2 viewed from the $y$-direction. The size distribution of the grains is taken from Li \& Draine (\cite{Li2001}). In the first frame the emission is assumed
to be directly proportional to the local gas density, while in the second frame
the actual calculated dust emission is used. These plots correspond to the
frequency of 353\,GHz (850$\mu$m). In the third frame the true
three-dimensional distribution of dust emission is also used, but the calculations
are done for 1500 GHz (200$\mu$m). In all cases a uniform value of $R=1$ was
assumed.
}
\label{nlambda}
\end{figure*}

In Fig.~\ref{nlambda} we have calculate the polarization for Core 2 under
different assumptions. In the first frame we use the Padoan et
al.~\cite{paal01} approach, and assume that the dust emission is directly
proportional to the density. In practice, this means that the weighting function
$f$ in Eq.~\ref{eqq} is simply the local density. As a further consequence,
the selected frequency has no effect on the polarization and affects 
only the intensity. The other two plots are calculated by taking
into account the actual three-dimensional distribution of dust emission, i.e.,
the weighting function $f$ is the local emission at the frequency in question.
On the other hand, in these plots the polarization reduction factor $R$ was
assumed to be constant, $R=1$. There is a clear, systematic change whereby the
dynamic scale of the intensity axis becomes more narrow.  The overall shape of
the figures remains rather similar, and there is only a small reduction in the
expected polarization degree.  The figures do, however, indicate that a change
in the wavelength affects the relative weights that are given to different
regions. When one moves to shorter wavelengths, the correlation between column
density and observed intensity decreases. Optically thick regions contain cold
dust and, therefore, radiate mainly at longer wavelengths. If one goes down
to far-infrared ($\sim 200\mu$m or below), the emission comes preferentially from
optically thin regions and from the surface of optically thick cores and filaments.
Therefore, shorter wavelengths work in a similar way as the variable
polarization reduction factor, and the observed polarization map contains less
and less information about magnetic fields inside the cores. Thus, the dense
cores become indistinguishable from their surroundings.

\subsection{Predictions for the Planck satellite}

Planck\footnote{http://www.rssd.esa.int/Planck/} is a CMB mission of the European Space Agency (ESA) that is planned to be
launched in 2007. The satellite will provide sensitive all sky maps in a wide
frequency range between 30 and 857\,GHz. In the shortest wavelength channels
the Galactic dust emission will be the dominant emission component. Planck has
polarization capabilities in seven channels (see Table~\ref{table:planck}). In
the following we examine the observable dust polarization at two frequencies,
143\,GHz and 353\,GHz. We use monochromatic simulations and no convolution is
done over the band profiles. However, the frequencies studied are relatively
far in the Rayleigh-Jeans regime, so that this simplification is not expected
to have a large effect, apart from a possible small error in the estimated
absolute values of the intensities. Our previous tests showed already that
morphological differences will arise only in comparison with much shorter
wavelengths (see Fig.~\ref{nlambda}).

\begin{table*}
\caption{%
The performance goal values for the resolution and sensitivity of the Planck
satellite. The expected sensitivities after 14 months of observations (two
full sky surveys) are given as thermodynamic temperature relative to the CMB
temperature of 2.73\,K. A more complete table is available through the ESA website
{\tt http://www.rssd.esa.int/?project=PLANCK\&page=perf\_top}
}
\label{table:planck} \centering
\begin{tabular}{ccccccc}
\hline 
\hline
Centre frequency   & Bandwidth  & Beamsize &  Sensitivity, $I$ & Sensitivity, $Q$ and $U$ & Sensitivity, $I$ & Sensitivity, $Q$ and $U$ \\
 (GHz) & (GHz)   & FWHM (arcmin) & [$\Delta$T/T$\times 10^{-6}$]  &
  [$\Delta$T/T$\times 10^{-6}$] & [$10^3$ Jy/sr] & [$10^3$ Jy/sr]\\
\hline
 30      &   6      &  33      &  2    &   2.8 &  0.15 &  0.21 \\
 44      &   8.8    &  24      &  2.7  &   3.9 &  0.42 &  0.60 \\
 70      &   14     &  14      &  4.7  &   6.7 &  1.7  &  2.4  \\
 100     &   33     &  9.5     &  2.5  &   4.0 &  1.6  &  2.6  \\
 143     &   47     &  7.1     &  2.2  &   4.2 &  2.3  &  4.4  \\
 217     &   72     &  5       &  4.8  &   9.8 &  6.3  & 13    \\
 353     &  116     &  5       &  14.7 &  29.8 & 12    & 24    \\
 545     &  180     &  5       &  147  &   -   & 23    & -     \\
 857     &  283     &  5       & 6700  &   -   & 26    & -     \\
\end{tabular}
\end{table*}

For this test we use model $C$. The model was previously scaled to represent a
cloud with a size of 6\,pc and mean density 640\,cm$^{-3}$, resulting in an
optically thick cloud with average visual extinction $<A_{\rm V}>\sim 6.3^{\rm
m}$. This was done in order to bring out the effects associated with high
densities and high values of the effective extinction (e.g., in
Eq.~\ref{eqaalg}). Here we want to study whether polarized signal might still
be detectable in regions of low column density. Therefore, we scale the column
densities down by a factor of ten, resulting in a model with average
extinction $<A_{\rm V}>\sim 0.6^{\rm m}$.  Simulations are done using a
variable polarization reduction factor $R$ that corresponds to the Weingartner
\& Draine (\cite{wedr01}) dust model. Noise is added to the total intensity
and the Q- and U-components according to the values given in
Table~\ref{table:planck}.

\begin{figure}
\resizebox{\hsize}{!}{\includegraphics{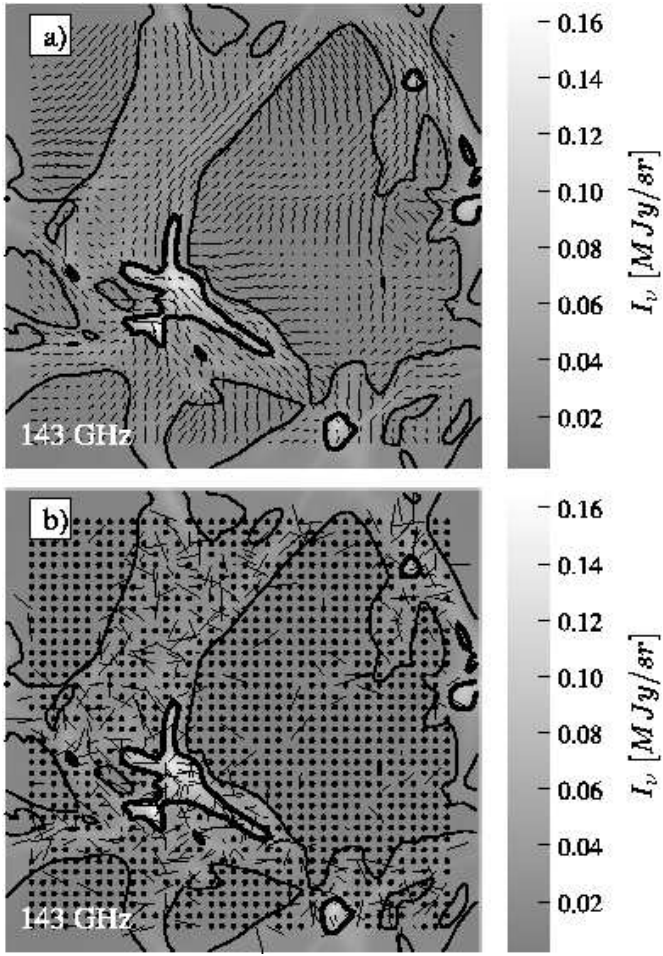}}
\caption{
Simulated polarization map at 143\,GHz without noise (upper frame) and with
noise corresponding to observations with the Planck satellite (lower frame). The cloud is
model $C$, scaled down to $<A_{\rm V}>$=0.6$^{\rm m}$. One pixel
corresponds to 7 arc minutes, the Planck resolution at this frequency.
Polarization vectors have been drawn for every third pixel, and the scaling of
the vectors is the same in both frames. In the upper frame the maximum
polarization degree is 9\%. In the bottom frame noise dominates and, for
clarity, long vectors are cut at 25\% and replaced by a dot. The background image
shows the total intensity, and the contours denote $A_{\rm V}$ values of 0.5
and 2.5 magnitudes.
}
\label{fig:planck_143}
\end{figure}

Figure~\ref{fig:planck_143} shows the computed polarization map for 143\,GHz.
In the figure we have also plotted contours corresponding to $A_{\rm V}$
values of 0.5 and 2.5 magnitudes. Below $A_{\rm V}\sim 0.5$ the polarized
signal is dominated by noise, which randomizes the angles and increases the
average polarization degree. Only above $A_{\rm V}\sim$2.5$^{\rm m}$ is
the signal comparable to the noise. On the other hand, for total intensity the signal to noise ratio is high and the noise has no visible effect on that parameter. The average intensity over this map was 1.8$\times 10^4$\,Jy/sr while the $1-\sigma$ noise, 2.3$\times 10^3$\,Jy/sr, is smaller about one order of magnitude.

\begin{figure}
\resizebox{\hsize}{!}{\includegraphics{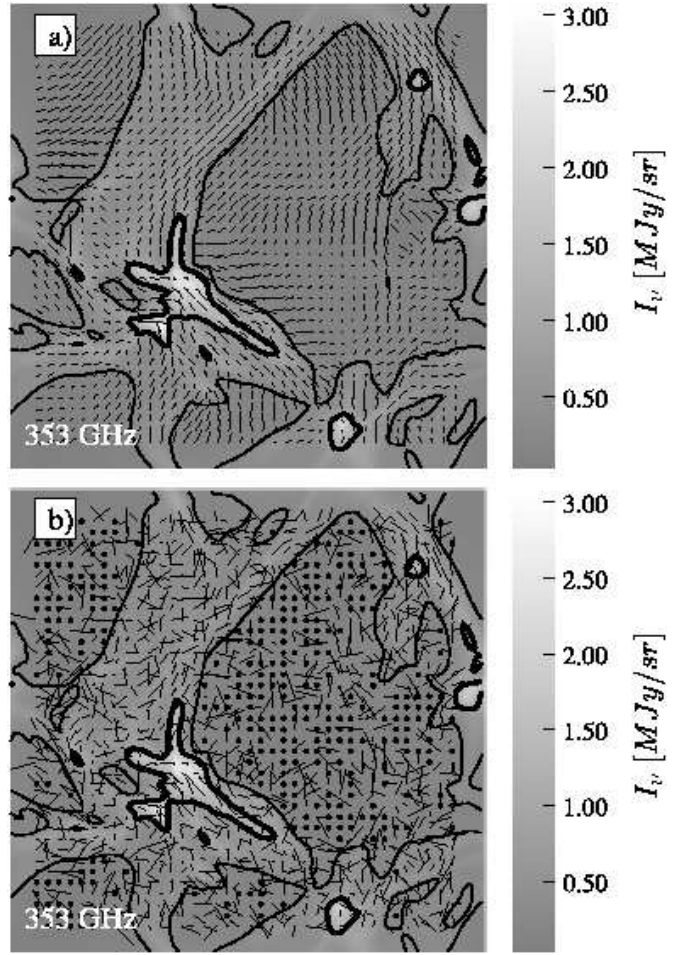}}
\caption{
Simulated polarization map at 353\,GHz without noise (upper frame) and with
noise corresponding to observations with the Planck satellite (bottom frame). 
Without noise the maximum polarization degree is 9\%. In the upper frame map, 
long vectors are cut at 25\% and replaced by a dot. One pixel corresponds to 
the Planck resolution of 5 arc minutes. Polarization vectors are drawn for every 
third pixel along both axes.
}
\label{fig:planck_353}
\end{figure}

In Fig.~\ref{fig:planck_353} a similar plot is shown for 353\,GHz. In terms of
absolute intensity, the sensitivity is lower at this frequency. However, the
rise in the dust spectral energy distribution more than compensates this, 
and the signal-to-noise
ratio is significantly higher.  The polarization vectors follow, with some
scatter, the magnetic fields even along filaments with $A_{\rm V}\sim 1^{\rm
m}$ or more. At 353\,GHz the spatial resolution of Planck is 5 arcmin so that
the pixel size is somewhat smaller than in Fig.~\ref{fig:planck_143}.

\begin{figure}
\resizebox{8cm}{10cm}{\includegraphics{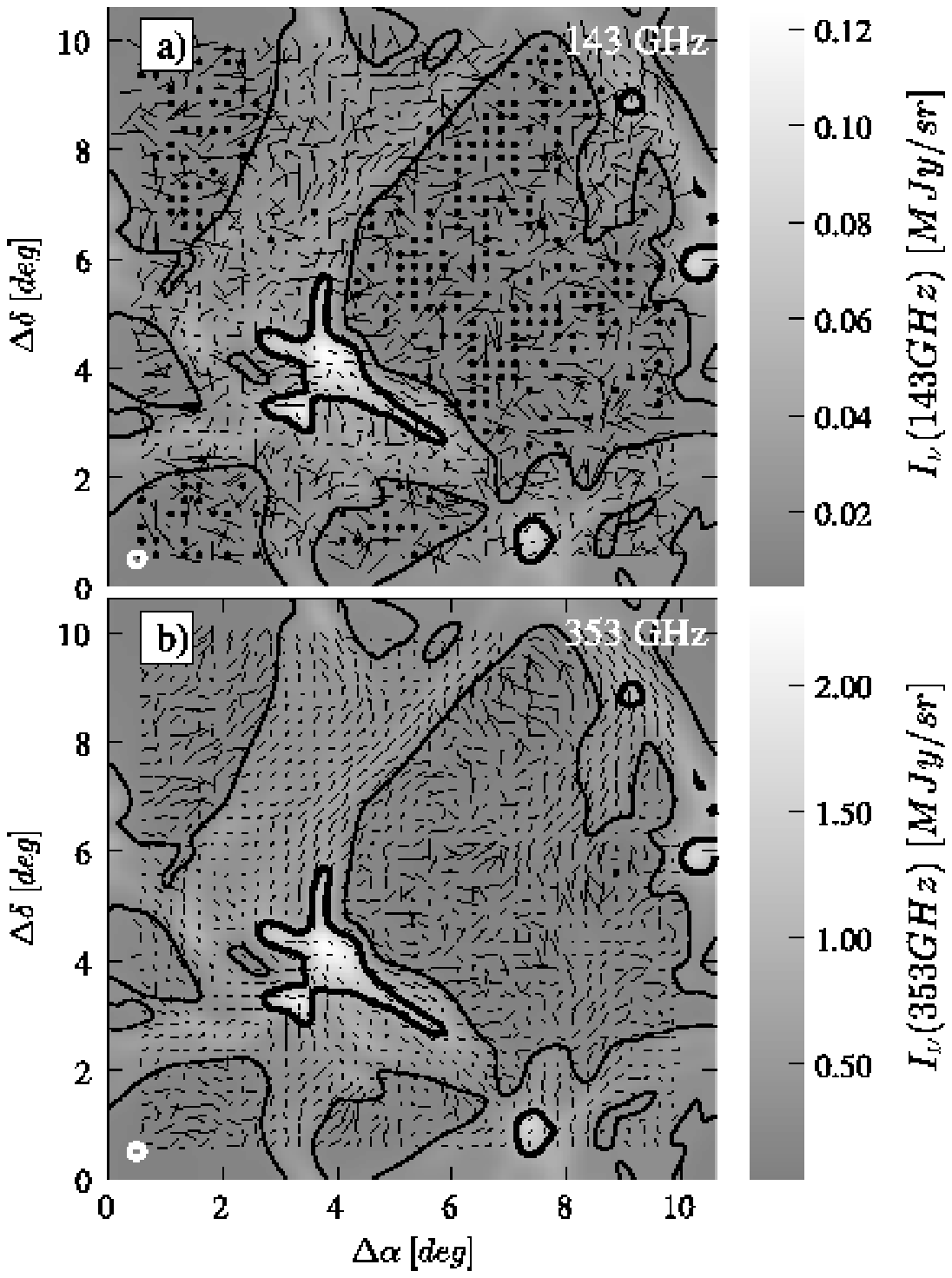}}
\caption{
The effect of convolution on the polarization maps shown in Fig.~\ref{fig:planck_143}
and \ref{fig:planck_353}. One pixel of the original simulation is assumed to correspond
to 5 arc minutes, and the maps are convolved down to a resolution of 15 arc minutes.
}
\label{fig:planck_fwhm3}
\end{figure}

The sensitivity can be increased by reducing the spatial
resolution. In Fig.~\ref{fig:planck_fwhm3} the pixel size of the previous 
maps is set equal to 5 arcmin, and the maps have been convolved down to a 
resolution of $FWHM=15$\,arc minutes. The signal-to-noise ratio is, of course,
correspondingly larger by a factor of $\sim 3$. At 353\,GHz a comparison with
Fig.~\ref{fig:planck_353}a shows that the orientation of the polarization
vectors is generally correct above $A_{\rm V}\sim 1^{\rm m}$, and is no longer
completely random even in the regions with the lowest column density. At 143\,GHz the scatter is larger, and thus even in the regions with $A_{\rm V} \sim 2.5^{\rm m}$ the true polarization field cannot be reliably measured.

\begin{figure}
\resizebox{\hsize}{!}{\includegraphics{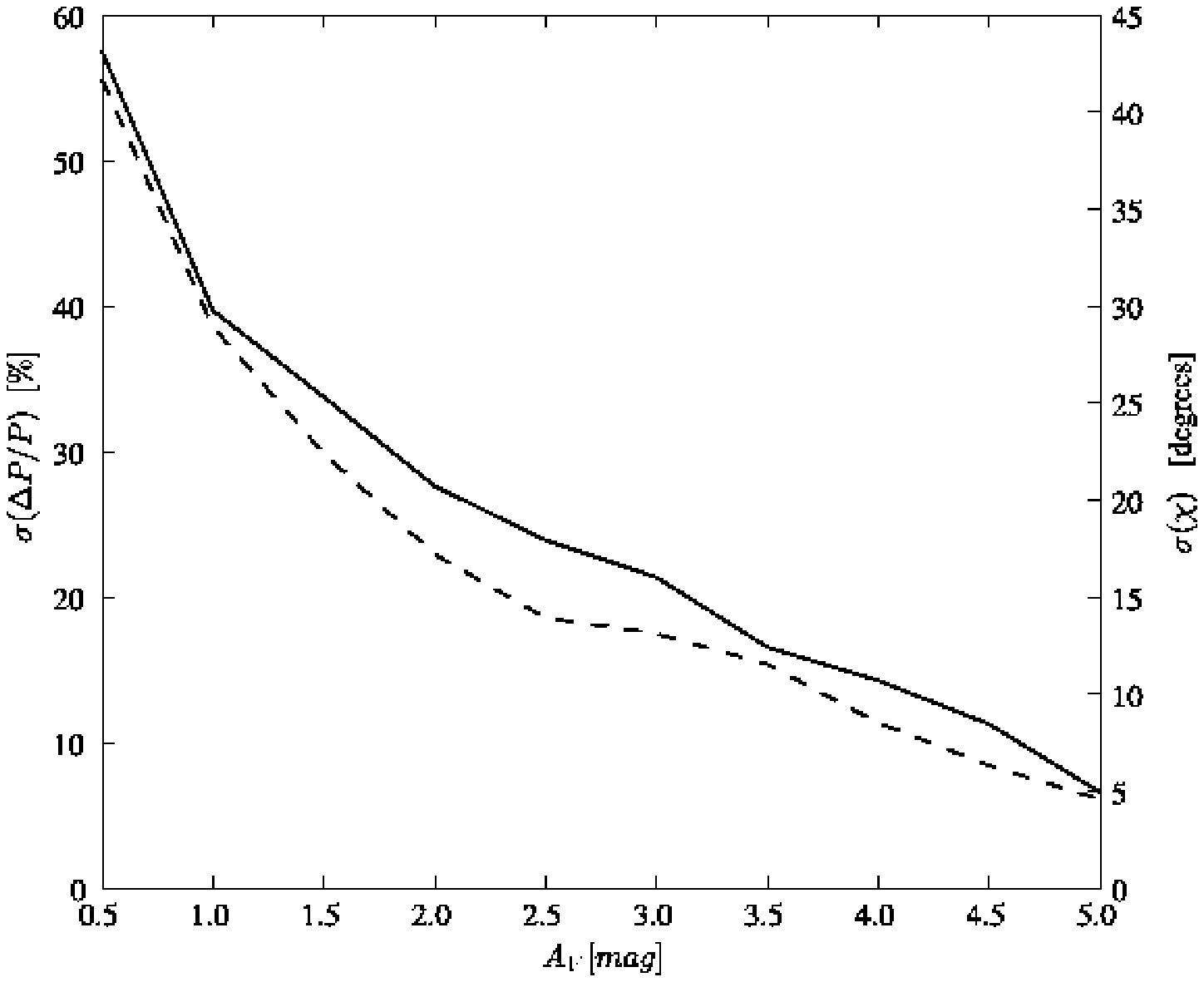}}
\caption{
The errors of polarization degree P (solid line, left hand scale) and polarization angle $\chi$ (dashed line, right hand scale) of the convolved 353\,GHz map in Fig.~\ref{fig:planck_fwhm3}.
}
\label{fig:planck_noise}
\end{figure}

The errors of polarization degree P (solid line, left hand scale) and polarization angle $\chi$ (dashed line, right hand scale) of the convolved 353\,GHz map are shown in Fig.~\ref{fig:planck_noise}. At $A_{\rm V}\sim 1^{\rm m}$ the relative error of P is $\sim$40\% and $\chi$ $\sim$30 degrees. At $A_{\rm V}\sim 2^{\rm m}$ the errors are already relatively low, $\sim$25\% for P and $\sim$15 degrees for $\chi$.

The 143\,GHz band is already close to the peak of the CMB spectrum but, even
at these column densities, the total signal is dominated by the dust emission.
In the previous 143\,GHz map the minimum column density is 1.1$\times
10^{20}$cm$^{-3}$, corresponding to less than 0.1 magnitudes of visual
extinction and a surface brightness of $\sim 3\times 10^{3}$\,Jy/sr. At this
frequency, the intensity of the CMB signal, $3.8\times 10^{2}$\,Jy/sr, is
still lower by almost one order of magnitude.  Because the polarization degree
is of same order for both the dust emission and the CMB, also the polarized signal
will be dominated by the dust emission. Conversely, the CMB signal introduces
errors to the estimated dust polarization signal only at the 10\% level. If one
wishes to study the dust emission at any fainter level, the CMB should be taken
into account as a possible error source. Of course, at higher frequencies the
contribution of the CMB signal becomes negligible, and the combination of
several wavelengths makes dust polarization studies possible even at fainter
levels, if the resolution is correspondingly decreased. Conversely, in Planck CMB observations, the dust contamination is minimized by using measurements at high galactic latitudes and lower frequency bands, 70\,GHz and 100\,GHz.

\section{Discussion}

\subsection{Current polarization model}

We have shown dust polarization maps computed for model clouds that are based
on three-dimensional MHD simulations. The main improvement with respect to
the previous study by Padoan et al. (\cite{paal01}) is that we have calculated
the true distribution of dust emission with the aid of three dimensional
radiative transfer calculations. This affects the way different volume
elements contribute to the observed signal. Furthermore, instead of resorting to
an ad hoc model, the polarization reduction factor has been calculated
according to the model of Cho \& Lazarian (\cite{cholazarian}). In this model
the grain alignment depends on radiative torques and, consequently, radiative
transfer calculations have also been used to estimate the strength of the
external radiation field at each position of the model clouds. This has been
accomplished through the determination of values of $A_{\rm V,eff}$. 

Observations show a decreasing polarization degree toward dense cores. Gon\c calves et al. (\cite{goal05}) proposed a geometrical effect, the toroidal magnetic fieldlines bending to resist the gravitational pull, to partly explain this drop. Padoan
et al. (\cite{paal01}) could explain the observed reduction by a qualitative model
where the dust alignment was restricted to regions with $A_{\rm V}$ below
3$^{\rm m}$. This extinction was calculated in similar fashion to our
$A_{\rm V, eff}$. With our models we find a very similar behaviour in the
plots of polarization degree vs. intensity (e.g., Fig.~\ref{cores}). At low
column densities the polarization degree varies between $\sim$4 and 10\%
while, at the centre of the cores, the degree drops to a few percent or, in
some cases, even below one per cent. In other words, our physical model is
capable of explaining the general observed behaviour (e.g., Ward-Thompson et
al. \cite{WT2000}; Henning et al. \cite{Henning2001}). At the same time,
the individual fields show considerable variations, and we find also cases
with no apparent decrease in the polarization degree (e.g., Core 3 and
direction $z$) or even increased polarization around intensity maximum (Core 3
and the filament seen toward direction $x$). This indicates that 
observations of individual cores may show considerable variation, depending on
the actual configuration of the magnetic field. One apparent problem is that
from observations it may be difficult to discern real and possibly
gravitationally bound cores from projection effects. In
particular, a core with large polarization reduction factor may appear very
similar to a filament with major axis and field lines parallel to the line of
sight.

The individual contributions of local density and $A_{\rm V, 1D}$ to the reduction of the polarization degree also merit a mention. In Model $C$ the local density varied between $\sim 1$ and $\sim 10^5\;{\rm cm^{-3}}$ while $A_{\rm V, 1D}$ had values from 0.005 to 4.3. Just by looking at Eq.~\ref{eqaalg} it is easy to see that in global scale, the variation in local density dominates over $A_{\rm V, 1D}$. However, in the cores, where the densities are around the same order of magnitude, the contribution of $A_{\rm V, 1D}$ comes to its own by emphasizing the difference between the illuminated regions on the outer parts of the core and the shadowed, even denser interior. In cores, both local density and $A_{\rm V, 1D}$ are of similar importance.

Related to this issue is the isotropic and anisotropic components of the radiation field, which play major part in the efficiency of grain alignment by radiative torques. In this work, we neglected detailed simulation of the anisotropy of the radiation field needed to derive the actual values of the anisotropy factor, $\gamma$, and instead used the fitting formula (Eq.~\ref{eqaalg}) by Cho and Lazarian (\cite{cholazarian}) where $\gamma = 0.7$. Draine and Weingartner (\cite{drwe96}) give $\gamma = 0.1$ for diffuse clouds, due to the anisotropy of the star distribution in the Galaxy. We can qualitatively estimate the effect that the change of $\gamma$ would have as it would be similar to the increase of local density by the corresponding inverse factor. For example, the difference between $10^2$ and $10^3\;{\rm cm^{-3}}$ at low $A_{\rm V, 1D}$ (see Fig.~\ref{Rfactors}) is a drop from 0.95 to 0.8 in $R$. Thus, we would expect a slightly lower polarization in those regions, but not by much. If anything, it would help to reduce the 'contamination' of the diffuse background and make the cores stand out more. Also, because we would expect to see a more isotropic field at the centre of the core than at its exterior, the inclusion of $\gamma$ would make the core centres drop in polarization even more sharply. In other words, our model is likely to overestimate the polarization at the core centres, rather than the opposite.

In the modelling we assumed dust properties to remain constant throughout the
model clouds, while only the alignment of the grains was dependent on the
local density and the visual extinction. However, there is ample evidence that
optical dust properties do change already above a few magnitudes of $A_{\rm
V}$. If this were caused primarily by a change in grain size distribution,
this could counteract the reduction obtained via the $R$ factor. As an
example, let us consider an optically thick core with density $n\sim
10^4$\,cm$^{-3}$ and $A_{\rm V, 1D}\sim 6^{\rm m}$. According to
Fig.~\ref{Rfactors} the $R$-factor could increase by a factor of several if
the L\&D grain distribution were replaced with a distribution with much
larger $a_{\rm max}$. On the other hand, Fig.~\ref{scatterR} shows that, in
the actual model, the effect of an increasing $a_{\rm max}$ was rather
limited. In that particular case, when L\&D dust model is replaced with MRN
type distribution with $a_{\rm max}\sim 1\mu$m, the polarization degree
increases in the core centre only from $\sim$2\% to $\sim$3\%. It is very
unlikely that in such a core $a_{\rm max}$ could grow close to one micron.
Another explanation for the relative insensitivity to $a_{\rm amax}$ is that
the observed intensity is weighted toward low density material. There are two
reasons for this. First, the polarization reduction factor decreases rapidly,
whatever the value of $a_{\rm max}$. Second, in a dense core the dust is
colder and, therefore, the emission is lower for a given column density. In a
core with $A_{\rm V,1D}\sim 7^{\rm m}$ to the centre, the dust temperature may already be
reduced to $\sim 14$\,K or below (Stamatellos \& Whitworth \cite{stwh03}). This can be compared with $\sim$18\,K in
diffuse material. In the Rayleigh-Jeans tail the effect is small, and even at
353\,GHz only a factor of 1.5. However, the effect becomes important when one
approaches the peak of dust emission. At 100$\mu$m the intensity produced by
an 18\,K grain is already ten times as large as the intensity produced by a
14\,K grain. From Fig.~\ref{Rfactors} one can see that at low densities the
slope of the $R$ vs. $A_{\rm V, 1D}$ relation is very similar for all values of
$a_{\rm max}$, and the percentile difference in the absolute value of $R$ are
similarly small. 

The above picture may change if dust grains undergo more profound changes. The
sub-mm emissivity of dust has been observed to increase significantly already
in clouds with visual extinction only a few magnitudes (Bernard et al.
\cite{bernard1999}, Stepnik et al. \cite{stepnik2003}). This has been
interpreted as evidence of not only grain growth but the appearance of dust
aggregates and fluffy grains of low packing density. At long wavelengths the
high emissivity might again increase the contribution that the densest regions
have on the total intensity observed towards cores. On the other hand, the
formation of dust aggregates can also affect the intrinsic polarization
efficiency of the grains and change the efficiency of the alignment mechanisms
(Lazarian \& Efroimsky \cite{Lazarian1999}; Gupta et al. \cite{Gupta2006}).
Such effects are not considered in this paper.

\subsection{Comparison with other simulations}

On larger scales, Prunet et al. (\cite{Prunet1998}) presented simulations of
polarized dust emission at high Galactic latitudes. The calculations were
based on three-dimensional cloud structure that was estimated based on HI
data. The HI line velocity provided information on the line-of-sight density
distribution. The simulations assumed constant dust properties with intrinsic
polarization efficiency $\sim$30\%. Furthermore, three field geometries were
considered. A uniform field direction resulted in high polarization degree
with very narrow distribution, and was dismissed as being incompatible with
observations. In our simulations this corresponds closely to model $A$, where
the field lines have remained almost parallel because of the strength of the
magnetic field. Of course, strong polarization also requires that the cloud is
not observed parallel to the field lines. In our model $A$, when the
sightlines were perpendicular to the field lines, the polarization degree
peaked around 10\%. For a random orientation the mean polarization would be
lower by half. In the other two cases studied by Prunet et al. the field was
assumed to be either parallel with the major axis of the cloud structures, or
to have a random orientation perpendicular to it. Both resulted in a wider
distribution of polarization degrees that, in the case of $R=1$, peaked around
2-3\%. These are again similar to our models $B$ and $C$, where, irrespective
of the viewing angle, we have an approximately log-normal distribution of
polarization degrees that peaks around the value of $\sim$3\%. Therefore,
although the results were obtained by entirely different methods, the basic
properties of the dust polarization seem very similar. This is reassuring for
attempts to build statistically meaningful all-sky templates for the polarized
dust emission that will be observed, e.g., by the Planck satellite. 

\begin{figure}
\resizebox{\hsize}{!}{\includegraphics{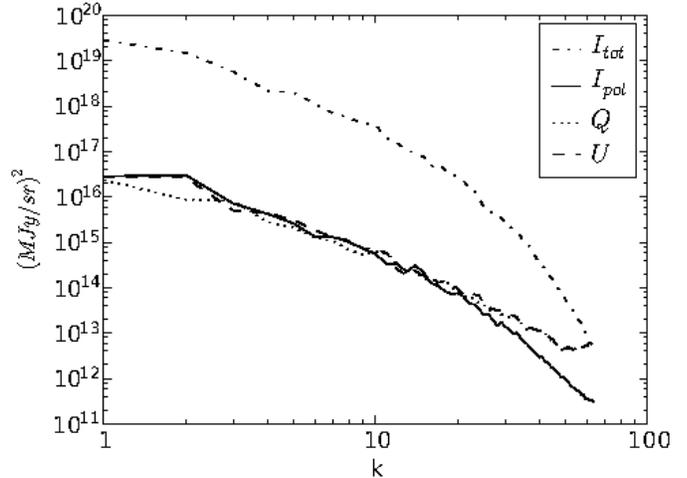}}
\caption{
The power spectra of the polarized intensity ${\rm I_{pol}}$ and the Stokes parameters ${\rm I_{tot}}$, Q and U are plotted for model $C$ from direction $y$. Column density $N$ follows the intensity power spectrum closely, and is thus not shown. While model $A$ showed a distinct shift in the level of the Q and U power spectrum in direction $z$, it - and model $B$ - were not different enough in shape to justify another figure.
}
\label{fig:powerspectra}
\end{figure}

Another important statistic is the power spectrum of the polarized signal.
Figure~\ref{fig:powerspectra} shows the power spectra computed for the model
$C$. Spectra are shown separately for the I, Q, and U components, as well as the polarized intensity ${\rm I_{pol}}$, from direction $y$. These can be compared with the power spectra presented by Prunet et al. (\cite{Prunet1998}). The slope of intensity, in interval $2 \leq k \leq 20$, is about $k^{-3}$, in agreement with Prunet et al. (\cite{Prunet1998}). The Stokes parameters Q and U have a slightly more shallow slope, but not as flat as by Prunet et al. (\cite{Prunet1998}). It can also be seen that in small scales, the power in the polarized components drops slower than the total intensity, making the polarized power an appreciable fraction of the total intensity power. The reason for this would appear to be the magnetic field morphology causing abrupt, pixel to pixel variations or larger structures, to about 10-pixel level, while the background intensity stays about the same.

\subsection{Comparison with observations}

Direct observations of polarized dust emission are still very scarce for all
but the very densest regions. The first detection of diffuse polarized dust
emission was reported by Beno\^it et al. (\cite{Benoit2004}). The observations
were done with the balloon-borne Archeops experiment, with instruments that
closely resemble those of the Planck satellite. Polarized signal could be
detected at 353\,GHz with the resolution of 13 arcmin, the average 
polarization degree being of the order of 4-5\%. However, a large variation
was seen between different regions. In the Cygnus region the polarization was
low, as can be expected if the magnetic field is preferentially parallel to
the spiral arms. On the other hand, in some regions of the Galactic plane the
polarization reached values of $\sim$15\%. In our models, polarization degrees this high could be reached only in model $A$ and only when the polarization reduction degree $R$ was equal to one (see Fig.~\ref{ld_poldeg_353}). However, our maximum polarization degree is somewhat arbitrarily set by our choice of $\alpha$, or to be more exact, our choice of $C_{\rm pol}/C_{\rm ran} = 0.15$. This ratio scales our maximum polarization degree to about 15.3\%, which explains why we do not reach $P = 15\%$ in all our models, since that would demand a total alignment and no depolarization on the line of sight. A slightly larger, yet still a moderate deviation from sphericity, can double our polarization degrees: at axial ratio of 1.2, $C_{\rm pol}/C_{\rm ran}= 0.3$. Thus one needs to be very careful not to overstate the case. Certainly a high polarization degree hints to an ordered magnetic field and high polarization efficiency throughout the cloud, yet even small changes in our assumptions about the grain shapes or lowering the density of our very dense model clouds would change the maximum polarization degree so that the other two models could duplicate even the high polarization degree of 15\%.

Ponthieu et al. (\cite{poal05}) presented the measurements of temperature and angular power spectra of the diffuse emission of Galactic dust at 353\,GHz as seen by Archeops on 20\% of the sky. They concluded that if the fraction of the sky observed by Archeops is representative, dust polarized radiation will be a major foreground for determining the polarization power spectra of the CMB at high frequencies above 100\,GHz. However, their power spectra cannot be directly compared with our results, because they cover much larger area, where the turbulence is no longer the dominant factor affecting the structure of the density and the magnetic fields.

\subsection{Future}

In Fig.~\ref{fig:planck_fwhm3} it is shown that when the spatial resolution of the Planck satellite is reduced to 15 arcmin, it will be able to provide a relatively clean map of the polarization of a diffuse cloud at 353\,GHz. The general morphology of the magnetic field is more or less correct above $A_{\rm V} \sim 0.5^{\rm m}$, even if the polarization degree P itself has a relative error of $\sim$60\%, as seen in Fig.~\ref{fig:planck_noise}. The situation improves with increasing $A_{\rm V}$, until at $A_{\rm V} \sim 2^{\rm m}$, the errors have dropped to the relatively low values of $\sim$25\% for P and $\sim$15 degrees for the polarization angle $\chi$. Thus, we conclude that Planck will be able to trace the general morphology of the magnetic fields at $A_{\rm V} \sim 1^{\rm m}$ and to map dust polarization reliably when $A_{\rm V}$ exceeds $\sim 2^{\rm m}$.

In this study we have used simulations with relatively low spatial resolution.
The results remain reliable at large scales, and the discussion of core
properties is appropriate for general density enhancements and the initial
phases in the formation of gravitationally bound cores. In a rotating and
collapsing core the magnetic field lines become wound so that both the field
strength and the depolarization caused by the field geometry increase. If the
system were not perfectly symmetric one might, at high resolution, detect
sharp variation in both the strength and direction of the polarization
vectors.  However, in order to be able to follow the development of magnetic
fields inside such collapsing objects, simulations with considerably higher
spatial resolution are needed. Spatially variable $\gamma$ and the inclusion of other grain alignment mechanisms would allow a more realistic polarization model. The work presented in this paper will be continued with such more detailed studies.

\begin{acknowledgements}

V.-M.P. and M.J. acknowledge the support of the Academy of Finland Grants no.
206049 and 107701. P.P. was partially supported by the NASA ATP grant
NNG056601G and the NSF grant AST-0507768.

\end{acknowledgements}

\end{document}